\documentclass{aa}  

\usepackage{url}
\usepackage[breaklinks,colorlinks,citecolor=blue]{hyperref}
\usepackage{graphicx}
\usepackage{txfonts}
\usepackage{lipsum}
\usepackage{subcaption}         
\usepackage{lscape}             
\usepackage{placeins}           

\newcommand{\tms}[2]{$#1\,\text{#2}$}
\newcommand{\tml}[2]{$t=#1\,\text{#2}$}
\newcommand{\tcc}[2]{$t_{\text{cc}}=#1\,\text{#2}$}
\newcommand{\ntrh}[3]{$n_{h}(#1\,\text{#2})=#3$}
\newcommand{\nh}[1]{$n_{h}=#1$}
\newcommand{\aplus}[1]{$A^{+}_{#1}$}
\newcommand{\figwidth}{0.9}

\begin{document}
   \title{Phase-space mixing of multiple stellar populations in globular clusters}
   
   \author{Francisco I. Aros\inst{1,2}\thanks{E-mail: francisco.aros@univie.ac.at},
           Enrico Vesperini\inst{2} \and Emanuele Dalessandro\inst{3}.}
   \institute{Department of Astrophysics, University of Vienna, T\"urkenschanzstrasse 17, 1180 Vienna, Austria
   \and
   Department of Astronomy, Indiana University, Bloomington, Swain West, 727 E. 3rd Street, IN 47405, USA
   \and
   INAF - Astrophysics and Space Science Observatory Bologna, Via Gobetti 93/3 I-40129 Bologna, Italy}
      
   \date{Received September 30, 20XX}

    \abstract
   {Globular clusters (GCs) host multiple populations characterised by abundance variations in a number of light elements. In many cases, these populations also show spatial and/or kinematic differences, which vary in strength from cluster to cluster and tend to decrease with the clusters' dynamical ages.}
   {In this work, we aim to study the dynamical mixing of multiple populations and establish a link between the more theoretical aspects of the mixing process and various observational parameters that quantify differences between the populations’ spatial concentration and velocity anisotropy.} 
   {We follow the dynamical mixing of multiple populations in a set of numerical simulations through their distribution in the energy and angular momentum phase space and quantify the evolution of their degree of dynamical mixing.}
   {We present the degree of dynamical mixing traced by the intrinsic differences in the phase space distribution of the populations. We compare the differences in phase-space with three observable quantities that describe the degree of mixing in the structural and kinematic differences of the populations: $A^{+}$, commonly used in the literature for spatial differences; and we introduce two new parameters, $\Delta A_{\beta}$ that traces the difference in velocity anisotropy and $\sigma_{\text{Lz}}$, that traces the angular momentum distribution of stars.}
   {Our study provides new insights into the dynamics of phase space mixing of multiple populations in globular clusters. We show that differences between the 1P and the 2P observed in old clusters contain key information on the cluster’s dynamics and the 1P and the 2P spatial and kinematic properties set by the formation processes, but caution is necessary in using the strength of the present-day differences to quantitatively constrain those imprinted at the time of formation.}

   
   \titlerunning{Multiple population mixing}
   \authorrunning{Aros, Vesperini \& Dalessandro}
   
   \maketitle
   
   \nolinenumbers

\section{Introduction}

Most globular clusters (GCs) show star-by-star variations in element abundances that point to distinct stellar populations. In general, observations show a first population (1P) with element abundances that are consistent with those of field stars having the same metallicity, and a second (2P) or more populations that present anomalous abundances in light elements (such as Na, O, Al, Mg, C, N, and He) while, in most cases, not showing any significant difference in heavy elements \citep[see,][ for reviews discussing evidence of multiple populations in spectroscopic and photometric studies]{gratton_2012,gratton_2019,bastian_2018,milone_2022}\footnote{See these reviews also for discussion of observational studies of a subset of about 20 per cent of the Galactic globular clusters showing evidence of a variation also in Fe.}.

The multiple stellar populations in GCs do not show only chemical differences but also differences in their spatial and kinematic properties: 2P stars are found to be spatially more centrally concentrated, and have internal kinematics characterized by a more rapid rotation and a radially anisotropic velocity distribution while the 1P population is more isotropic or tangentially anisotropic with the strength of these dynamical differences found to decrease with the cluster's dynamical age \citep[see, e.g. ][ and references therein for studies of the spatial and kinematic differences of multiple stellar populations in GCs including examples of dynamically older clusters where differences between the 1P and the 2P dynamical properties have been erased by the GC dynamical processes]{dalessandro_2019,dalessandro_2024,libralato_2023,cordoni_2020,cordoni_2025}. The dynamical differences revealed by observations can provide some insight into the initial dynamical state of the populations emerging from the formation process, but investigations of the dynamics of multiple populations are necessary to establish a closer and more quantitative link between the initial and present-day spatial and kinematic properties as evolutionary processes can lead to different degrees of dynamical mixing during a cluster’s evolution.

Various formation scenarios have been proposed for multiple populations in GCs \citep[see, e.g.,][for a few reviews]{bastian_2018,gratton_2019,milone_2022}. While there is no consensus on many aspects of the formation of multiple populations, most models agree that the 2P would form more centrally concentrated than the 1P \citep[see, e.g.][]{dercole_2008,bekki_2010,bekki_2011,gieles_2018,calura_2019,wang_2020,lacchin_2022}. In the last decade, several works have studied the long-term evolution of GCs with multiple populations through simulations, focusing on the effects of dynamical processes on the evolution of the structural and kinematic properties of multiple populations \citep[see, e.g.][]{bekki_2010,mastrobuono-battisti_2013,vesperini_2013,vesperini_2018,vesperini_2021,henault-brunet_2015,tiongco_2019,sollima_2021,lacchin_2022,hypki_2024}. 
Some investigations have also explored the possibility that some of the clusters with more complex chemical properties characterised by a significant spread in iron may result from the mergers of clusters with different iron abundances [see, e.g., \cite{amaro_seoane_2013,gavagnin_2016,khoperskov_2018}; see also \cite{ishchenko_2023} for a study of the possible history of Galactic globular cluster collisions].

In this work, we follow the long-term evolution of GCs with multiple populations to characterise the dynamical mixing process in the energy-angular momentum phase space, $E-L$, and to connect the phase-space mixing process with observational tracers for the spatial and kinematical differences between the populations. We also explore the strength of the signatures of these dynamical differences over the dynamical history of the GCs and the possibility of establishing a link between the initial and the present-day dynamical properties of multiple populations.

The outline of the paper is the following. Section \ref{sec:sims} describes our simulations and the initial conditions of our models. In section \ref{sec:internal_mixing}, we explore the mixing process through the behaviour of the stellar distribution in the energy and angular momentum phase space of each population. In section \ref{sec:proj}, we analyse three observable properties that characterise multiple populations' spatial and kinematic differences, and we compare them with the mixing process in phase space. Finally, in section \ref{sec:summary}, we discuss and summarise our results.


\section{Initial conditions and simulations of Globular Clusters with multiple stellar populations.}
\label{sec:sims}

We follow the long-term evolution of seven simulated globular clusters with two distinct stellar populations. All models were evolved using the \texttt{MOCCA} code \citep{giersz_2013,hypki_2013}, which is based on the Henon’s Monte Carlo methods \citep{henon_1971a,henon_1971b} and includes prescriptions for stellar evolution \citep[through the \texttt{SSE} and \texttt{BSE} codes][]{hurley_2000,hurley_2002}, close stellar encounters \citep[through the \texttt{FEWBODY} code][]{fregeau_2004} and the dynamical formation of binaries. \texttt{MOCCA} accounts for the effects of a host galaxy tidal field by including a tidal radius truncation in the cluster \citep[][for further details on the \texttt{MOCCA} code]{hypki_2013,giersz_2013}. The time variation of the tidal fields associated with eccentric orbits or the tidal shocks due, for example, to Galactic disk crossings, are not included, and the study of those effects require \textit{N}-body simulations that properly account for a realistic modelling of the external tidal field \citep[see, e.g.,][]{cai_2016,berczik_2025}.

Five of the seven models presented here were also used in \cite{livernois_2024} to study energy equipartition in clusters with multiple populations. Here, we briefly describe the initial conditions of our models and refer to \cite{livernois_2024} for further details. The choice of initial conditions generally follows the results of simulations modelling the formation of multiple populations and finding that 2P stars form in a high-concentration system in the innermost regions of the 1P system \citep[see e.g.,][]{dercole_2008,bekki_2010,calura_2019}. All simulations start with the first population modelled as a spherical system following the density profile of a \cite{king_1966} model with a central dimensionless potential of $W_{0}=5$ and the second population modelled as a King model with $W_{0}=7$. All models start with a tidally filling 1P population extending to the cluster's tidal radius, while the 2P population is concentrated in the innermost regions of the 1P system with an initial ratio of the 1P to the 2P half-mass radii equal to 20 or 10 (see Table \ref{tab:sim_prop}). For all the simulations, we adopt an initial mass ratio between populations of $M_{\text{1P}}/M_{\text{2P}} = 4$. We evolve clusters with $2\times10^6$, $1\times10^6$ and $0.5\times10^6$ stars. All models have an initial tidal radius corresponding to that expected for clusters with the given mass at a Galactocentric distance of $4\,\text{kpc}$ in a logarithmic potential with a circular velocity of $220\,\text{km/s}$. We also explore the case for a
weaker tidal field (corresponding to a galactocentric distance of $8\,\text{kpc}$, for the model with $0.5\times10^6$ stars (N05M-wf in Table \ref{tab:sim_prop}).

The initial conditions adopted follow the general trend of different possible formation scenarios predicting a centrally concentrated 2P population and are able to produce general properties at $12\,\text{Gyr}$ consistent with those observed in GCs \citep[see, ][and discussion therein]{vesperini_2021,livernois_2024,hypki_2022}. All simulations start with stellar masses of 1P and 2P stars following a Kroupa IMF \citep{kroupa_2001} between $0.1$ and $100\,M_{\odot}$ and without primordial binaries.

The goal of this work is to study the dynamics behind the multiple population mixing in phase space; the initial conditions explored allowed us to address some specific points concerning the mixing process, specifically:

\begin{itemize}
    \item[(a)]{Dynamical age and relaxation time: The different models we have simulated reach diﬀerent dynamical ages and degrees of mixing at the end of the simulations, with the dynamical age increasing for decreasing initial masses. This allows us to characterise the phase space properties of systems at different stages of their mixing dynamical history.}
    \item[(b)]{Primordial velocity anisotropy: All models except the N1M-ra model start with an isotropic velocity distribution. The N1M-ra model starts with a primordial radial anisotropy following an Osipkov-Merrit velocity anisotropy profile \citep{osipkov_1979,merrit_1985}. This anisotropy profile defines the kinematics of the whole cluster without distinction between populations. The centrally concentrated 2P stars populate the central regions of the clusters where the velocity distribution is close to isotropic, while the regions characterised by the stronger radial anisotropy are initially dominated by the 1P stars.} 
    \item[(c)]{Different concentrations of the second population: The five models in common with \cite{livernois_2024} have an initial half-mass radius ratio between the populations of $r_{\text{h,1P}}/r_{\text{h,2P}} = 20$. We have included two new models with an initial concentration of $r_{\text{h,1P}}/r_{\text{h,2P}} = 10$ (named "r10"), which makes the second population less concentrated.}
\end{itemize}

Table \ref{tab:sim_prop} summarises the principal differences in initial conditions for the model analysed in this work. We save each model state at every \tms{100}{Myr} to have a continuous description of the time evolution of the properties we analyse here.  

\begin{table}
\caption{\label{tab:sim_prop}Initial conditions}
\centering
\begin{tabular}{l|c|c|c|c|l}
\hline\hline
Name  & N & $M_{\text{tot}}$ & $r_{\text{h,1P}}/r_{\text{h,2P}}$ & $r_{\text{tidal}}$ & $\beta_{t=0}$  \\
      & $(10^6)$ & $(10^6\,M_{\odot})$ &   & $(\text{pc})$ & \\

\hline
N2M      & $2.0$ & 1.273 & 20 & $97$ & iso \\
N1M      & $1.0$ & 0.637 & 20 & $77$ & iso \\
N1M-ra   & $1.0$ & 0.639 & 20 & $77$ & ani \\
N05M     & $0.5$ & 0.319 & 20 & $61$ & iso \\
N05M-wf  & $0.5$ & 0.319 & 20 & $97$ & iso \\
\hline
N2M-r10  & $2.0$ & 1.275 & 10 & $97$  & iso \\
N1M-r10  & $1.0$ & 0.638 & 10 & $77$  & iso \\  
\hline
\end{tabular}
\tablefoot{N is the total number of stars, $M_{\text{tot}}$ is the total mass of the cluster, $r_{\text{h,1P}}/r_{\text{h,2P}}$ is the initial ratio of the 1P to 2P half-mass radius, $r_{\text{tidal}}$ is the cluster tidal radius, and $\beta_{t=0}$ denotes initial velocity anisotropy  ("iso" for isotropic or "ani" for anisotropic). All simulations start with a ratio $r_{\text{h}}/r_{\text{tidal}}$ equal to about $0.14$ and $r_{\text{h,2P}}/r_{\text{tidal}}$ equal to about $0.01$ ($0.02$ for the N2M-r10 and the N1M-r10 simulations).}
\end{table}

For the subsequent analysis, we will measure a cluster's dynamical age with the ratio of the cluster's physical age to its half-mass relaxation time calculated at that age, $n_{h} = t/t_{rh}$. We use the Spitzer’s half-mass relaxation time \citep{spitzer_1987}: 
\begin{equation}
t_{rh} = \frac{6.5\times 10^2 \text{Myr}}{\ln(\lambda N)}\left({\frac{M}{10^5M_{\odot}}}\right)^{1/2}\left({\frac{M_{\odot}}{\langle m\rangle}}\right)\left({\frac{r_{h}}{\text{pc}}}\right)^{3/2}\,,    
\end{equation}
where $\lambda=0.02$ \citep{giersz_1996}, $N$ is the total number of stars in the cluster, $M$ the total mass of the cluster, $\langle m\rangle$ is the mean mass, and $r_h$ is the half-mass radius.

In Tables \ref{tab_app:sim_prop_t07} and \ref{tab_app:sim_prop_t12}, we report the properties of all our models at $7\,\text{Gyr}$ and $12\,\text{Gyr}$, respectively. The final properties are generally consistent with those of many Galactic GCs, but we strongly emphasise that these models are not meant to describe any specific observed GCs, nor do the initial conditions explored provide full coverage of the parameter space of possible initial and final conditions. The models are meant to provide a starting point to understand the mixing process of multiple populations and the current and new tools we are introducing to quantify the degree of mixing.    

\section{Signatures of phase space mixing.}
\label{sec:internal_mixing}

We study the degree of dynamical mixing by tracing the time evolution of the distribution function (DF) for each stellar population. The simulations we use are spherically symmetric, and therefore, we expect that the DF is only a function of the energy ($E$) and total angular momentum ($L=|\vec{L}|$). We identify the DF of each population as $F_{1P}(E,L)$ for the initially extended 1P population and $F_{2P}(E,L)$ for the initially compact 2P population. As the cluster evolves, both populations, which were initially in different regions of the phase-space of energy and angular momentum, will diffuse in phase space and eventually populate the same region; once this happens, we consider the populations to be fully mixed.

We follow the mixing of the stellar populations in each simulation by measuring the instantaneous energy and angular momentum per unit mass for each star every \tms{100}{Myr}. We measure these values by taking the three-dimensional positions and velocities of each star along with the potential energy given by the global mass distribution in the cluster:
\begin{align}
    & E_{i} =  0.5\times(v_{r,i}^2+v_{\theta,i}^2+v_{\phi,i}^2) +\Phi(r_{i})\label{eq: energy}\\
    & L_{i} =  r_{i}\times\sqrt{v_{\phi,i}^2+v_{\theta,i}^2}.\label{eq: angular_momentum}
\end{align}
To compare simulations with different sizes and masses, we normalise the energy by $\Phi_{h}=0.5GM_{\text{tot}}/r_{h}$ and the angular momentum by $L_{h}=\sqrt{0.5GM_{\text{tot}}r_{h}}$; that corresponds to the potential energy of a point mass of half of the total mass of the cluster at the distance of $r_{h}$, and the angular momentum of the circular orbit at $r_h$ in the same point mass potential.  

\begin{figure*}[h!]
    \centering
    \includegraphics[width=\figwidth\linewidth]{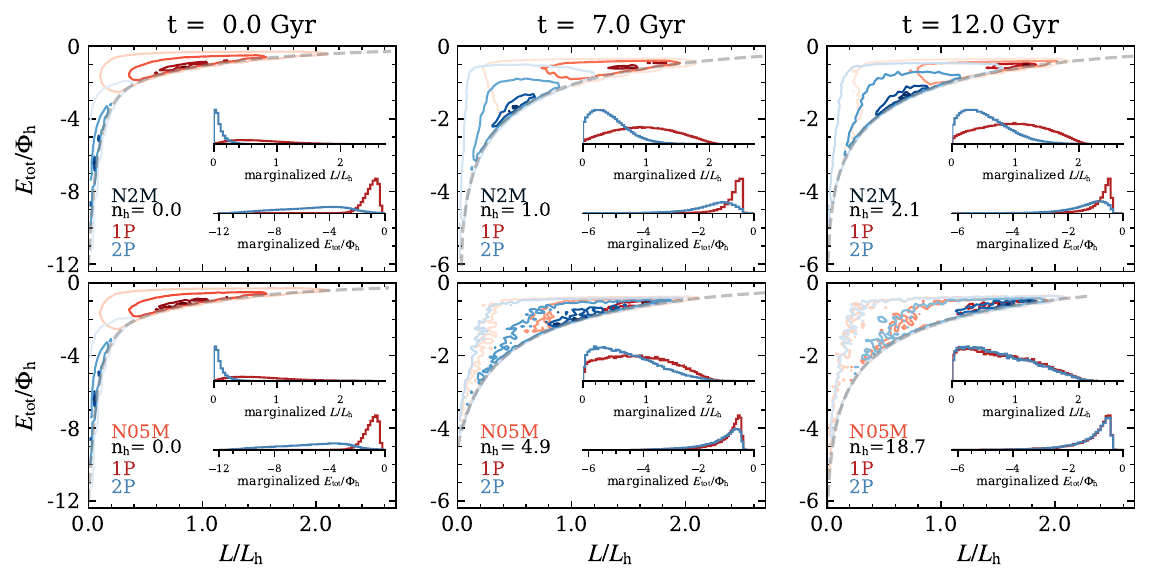}
    \caption{Phase space distribution of stars in the first and second populations for the models N2M and N05M at different times. At \tml{0}{Gyr}, both clusters show a similar relative distribution of stars in the first (red) and second (blue) populations due to having the same half-mass radius and mass ratios.  As the clusters evolve, the second population expands towards the regions initially dominated by the first population.  By \tml{12}{Gyr}, the N2M model is still partially mixed, while model N05M has a significant degree of mixing. The insets show the marginalised distribution of energy and angular momentum, also highlighting the evolution of dynamical mixing. In all panels, the dashed lines follow the maximum allowed angular momentum at a given energy, defined by the angular momentum of the circular orbit of the corresponding energy.}
    \label{fig:phase-space-mixing-example}
\end{figure*}

Figure \ref{fig:phase-space-mixing-example} shows the DF of each stellar population for the simulated clusters with the least (N2M) and most (N05M) degree of mixing at three different epochs: \tms{0}{Gyr}, \tms{7}{Gyr} and \tms{12}{Gyr}. As described in section \ref{sec:sims}, both clusters have different initial conditions but share the same relative ratios for each population's half-mass radius and total masses, giving them similar normalised distributions in the phase space. Initially ($t=0$), the first population (1P, red contours) is extended (larger values of $L$) and less concentrated (shallower total energy), and it occupies the upper regions of the phase space. The second population (2P, blue contours) is more concentrated and limited to the deepest part of the cluster’s potential. The inset-panels show the initial difference between the configurations of both populations in the marginalised distributions of angular momentum and total energy. As the clusters evolve, both populations start to mix in phase space. The second population “diffuses” to larger energies and higher angular momentum values, a process also driven by the reduction of the potential caused by the cluster’s mass loss due to stellar evolution and stellar escape. Note that the increase of the angular momentum values also goes in hand with the more extended orbits. At \tms{7}{Gyr} (middle panel), we observe that model N2M still has a significant difference between the two populations, while model N05M has a higher degree of mixing, particularly for the energy distribution. The N05M model is also dynamically older with a dynamical age of \ntrh{7}{Gyr}{4.9} compared with the model N2M with \ntrh{7}{Gyr}{1.0}. At \tms{12}{Gyr}, the populations in model N2M are still in the process of mixing. On the other hand, both populations appear thoroughly mixed for model N05M, as the distributions of stars only show minor differences. 

We construct a grid on the energy-angular momentum space to quantify the differences in the distribution of both stellar populations. In each bin, we take the squared difference of the normalised distributions and then sum over all bins: 
\begin{equation}
    S^2 = \sum_{i}^{N_{L}}\sum_{k}^{N_{E}}{\left(\rho(L_{i},E_{k})_{2P}-\rho(L_{i},E_{k})_{1P}\right)^2}\times (N_{L}N_{E})\,,
    \label{eq:squared-difference}
\end{equation}
where $\rho(L_{i},E_{k})_{1P}$ and $\rho(L_{i},E_{k})_{2P}$ are the number of stars in each $(i,k)$ bin normalized by the total number of stars in each population to make the distributions comparable. While the initial value of $S^2$ might depend on the initial separation of the populations, as the populations mix, $S^2$ tends to zero. 

\begin{figure}[h!]
    \centering
    \includegraphics[width=\figwidth\linewidth]{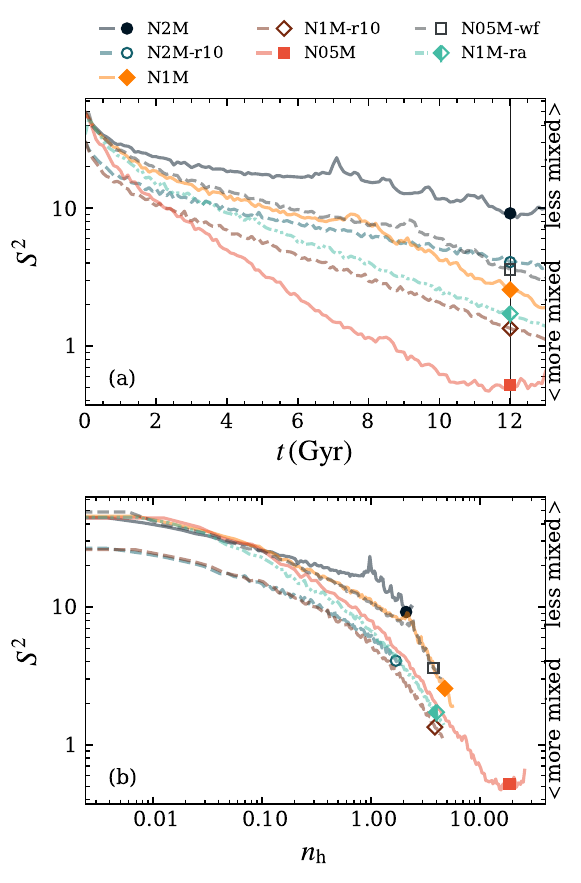}
    \caption{Evolution of the $S^2$ parameter for all models. Panel (a) shows the evolution of $S^2$ with respect to the physical time in the simulations. All models show different degrees of mixing at most times. At \tml{12}{Gyr} (vertical line), model N2M is the least mixed, while model N05M is the most mixed in phase space. Panel (b) shows the evolution of $S^2$ as a function of the number of half-mass relaxation times. The symbols show the value of $S^2$ at \tms{12}{Gyr} and the respective number of relaxation times.}
    \label{fig:s2-comparison}
\end{figure}

Figure \ref{fig:s2-comparison} shows the evolution of $S^2$ for all models in our sample as a function of time (panel (a)) and their dynamical age $n_{\text{h}}$ (panel (b)). At \tms{12}{Gyr}, the models show a clear diversity in the degree of mixing for the clusters. The model N2M is the least mixed one with $S^2 = 9.99$, while the model N05M shows the most mixing at \tms{12}{Gyr} with $S^2 = 0.52$. In between these two extremes, we find two groups with similar $S^2$, first the models N2M-r10 ($S^2=4.07$), N1M ($S^2=3.62$) and N05M-wf ($S^2=3.59$), and then the models N1M-r10 ($S^2=1.77$) and N1M-ra ($S^2=1.61$). All models follow a more similar evolutionary path of $S^2$ with respect to the dynamical age $n_h$ (panel (b)). One substantial difference arises for the N2M model where $S^2$ increases around \nh{1}, triggered by the cluster's core collapse, which we will discuss later in this section. The models N2M-r10 and N1M-r10 also follow a common evolutionary path; these models started with a less concentrated second population and are initially more mixed, which translates to a lower initial value of $S^2$. The N2M-r10 and N1M-r10 models follow a similar but parallel evolutionary path to the N2M, N1M and N05M models. The N05M and N1M-ra models start with comparable values of $S^2$ as the N2M, N1M and N05M-wf models, but the populations mix more rapidly.

\subsection{Core-collapse}
\label{sec:internal_mixing_cc}

\begin{figure}
    \centering
    \includegraphics[width=\figwidth\linewidth]{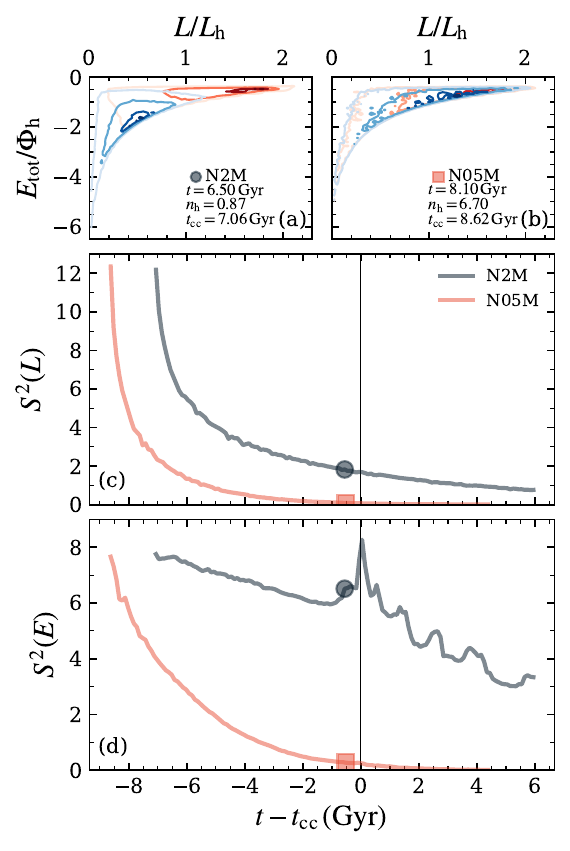}
    \caption{Effects of partial mixing during core collapse. Panels (a) and (b): phase space distribution of the models N2M and N05M at $\sim500\,\text{Myr}$ before core collapse. The N2M model is far from fully mixed, while the N05M model has a significant degree of mixing. Panel (c): evolution of the $S^2$ parameter centred at the time of core collapse for the models N2M and N05M (\tcc{7.06}{Gyr} and \tcc{8.62}{Gyr}), the markers indicate the corresponding time of the phase space distributions in panels (a) and (b). Panel (d) shows the value of the $S^2$ parameter only for the energy component of the phase space. The peaks in $S^2(E)$ show the effects of core collapse and subsequent core oscillations for the model N2M where the second population dominates the inner regions, while for N05M model, the two populations are almost completely mixed at the time of core collapse and the evolution of $S^2(E)$ does not show any feature associated to the core collapse.}
    \label{fig:s2-corecollapse}
\end{figure}

Four models in our sample undergo core collapse within the \tms{13}{Gyr} of evolution (see Fig \ref{fig-app:all_diff_3d}). From those, we see in Figure \ref{fig:s2-comparison} that the N2M model shows multiple peaks in $S^2$. While not as strong, the N1M, N05M, and N05M-wf models also show a peak in $S^2$ (panel (a) in Fig \ref{fig:s2-comparison}). These peaks are consistent with the times of the first core collapse. Models N1M-ra, N2M-r10 and N1M-r10 do not undergo core collapse within the $13\,\text{Gyr}$ of evolution.

We now discuss how the small peaks in $S^2$ found for some of the simulations are the fingerprints of the evolution of the innermost regions at the time of core collapse and how they are related to the degree of spatial mixing of the 1P and 2P populations in the cluster's central regions at that time.

The N2M model undergoes core collapse at \tml{7.06}{Gyr} (\ntrh{7.06}{Gyr}{0.97}). The two populations do not show a significant mixing in phase space at this time, where the second population dominates the cluster’s centre with a fraction of the total number of stars belonging to the 2P population within the 1 per cent Lagrangian radius equal to $0.87$ while the global 2P fraction is equal to 0.52. Panel (a) in Fig \ref{fig:s2-corecollapse} shows the phase space at $\sim$\tms{500}{Myr} before the first core collapse, showing that the populations are not mixed yet. On the other hand, the N05M model is considerably more mixed before it undergoes core collapse at \tml{8.62}{Gyr} (\ntrh{8.62}{Gyr}{7.46}), as shown in panel (b) of Fig \ref{fig:s2-corecollapse}. The N05M model has a central second population fraction of $f_{\text{2P}}(r_{1\%})=0.63$ and a global fraction of $f_{\text{2P}}=0.6$ at the time of the core collapse.

The difference in the central concentration of each population drives the observed small peaks in $S^2$ during the core collapse. Panel (c) in Fig \ref{fig:s2-corecollapse} shows the value of $S^2$ calculated using only the angular momentum distribution (i.e. marginalising over the energy space),$S^2(L)$, and that calculated using only the energy distribution, $S^2(E)$. In both simulations, the time evolution of $S^2(L)$ does not show any feature associated with core collapse, 
while $S^2(E)$ for the N2M simulation is characterized by the same peaks associated to core collapse and the subsequent core oscillations already revealed by the evolution of $S^2$ in Fig \ref{fig:s2-comparison}. As the core starts contracting, in systems like the N2M models where 2P stars are the dominant population in the inner regions, more second-population stars follow the core contraction than first-population stars; this affects the distribution of potential energies of the two populations and produces the observed peaks in $S^2$ and $S^2(E)$. The strength of the peaks is determined by the degree of mixing: the further the system is from mixing, the more prominent the peaks in $S^2$ and $S^2(E)$ are.

\subsection{Initially anisotropic model}
\label{sec:internal_mixing_ani}

The N1M and N1M-ra models start with the same initial conditions except for the initial radial anisotropy profile. The presence of primordial radial anisotropy impacts the overall evolution of the cluster. The N1M model has lost more mass and stars than the N1M-ra model; furthermore, the N1M model undergoes core collapse at \tml{7.54}{Gyr} while the model N1M-ra is still evolving towards core collapse at \tms{13}{Gyr}. The delay of the core collapse due to primordial velocity anisotropy has been thoroughly described by \cite{breen_2017} and \cite{pavlik_2021} in the case of single-populations clusters.

\begin{figure}
    \centering
    \includegraphics[width=\figwidth\linewidth]{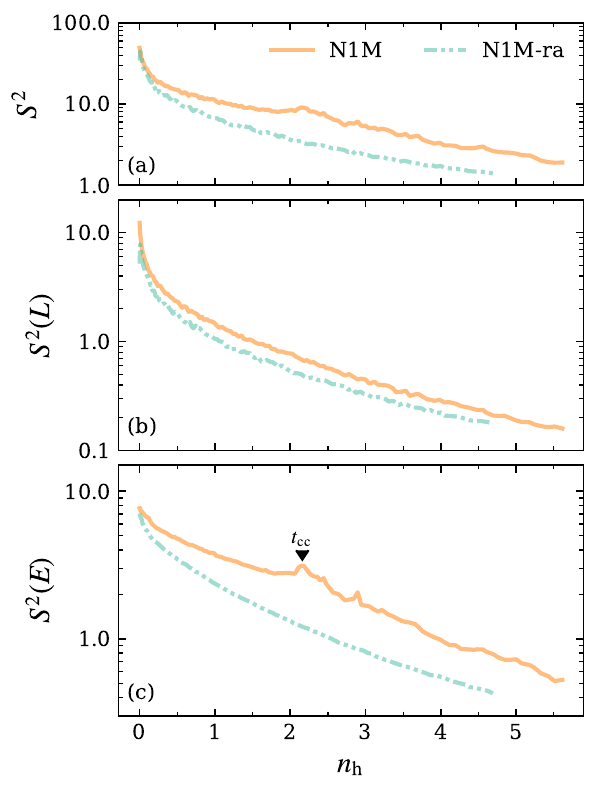}
    \caption{Evolution of the $S^2$ parameter (a), as well as $S^2$ measured over the angular momentum ($S^2(L)$, panel (b)) and energy ($S^2(E)$, panel (c)) spaces for the N1M and N1M-ra models. The primordial radial anisotropy increases the population mixing rate while keeping the cluster dynamically younger. The model N1M-ra does not undergo core collapse within its \tms{13}{Gyr} of evolution.}
    \label{fig:s2-anisotropy}
\end{figure}

Figure \ref{fig:s2-anisotropy} shows the $S^2$ parameter, as in Figure \ref{fig:s2-comparison}, for the N1M and N1M-ra models and the marginalised distribution differences: $S^2(L)$ and $S^2(E)$. The N1M-ra model develops more dynamical mixing than the N1M model within their common range of dynamical ages. Two effects drive the difference in mixing as accounted by the $S^2$ parameter. The initial velocity anisotropy in the N1M-ra model makes the angular momentum distribution for the first population more compact and similar in extent to the second population (see Figure \ref{fig-app:phase-space-all-B}). As the cluster evolves, the second population develops radial anisotropy while the first population slowly loses its anisotropy, becoming isotropic at about \nh{3} (see Figure \ref{fig:ani_areas_all} and discussion therein). On the other hand, the N1M model evolves quickly to have a radially anisotropic second population and a tangentially anisotropic first population, adding to the differences in the angular momentum distribution. The second effect is the delayed core collapse, which allows the two populations in the N1M-ra model to steadily spatially mix. In contrast, in the N1M, and as discussed in Section \ref{sec:internal_mixing_cc}, the core collapse delays the energy mixing if one of the two populations dominates the core.

\section{Observable population mixing parameters derived from simulations.}
\label{sec:proj}

We discussed in Section \ref{sec:internal_mixing} how differences in the phase-space distribution, traced by the $S^2$ parameter, can help quantify the degree of dynamical mixing of the two populations in a globular cluster. However, the $S^2$ parameter can be measured only in numerical simulations. In this section, we explore three observable tracers for multi-population mixing, and we compare them with the $S^2$ parameter. 

Stars considered for the analysis presented in this section are limited to those brighter than two magnitudes below the turn-off magnitude. This limit encompasses the range of current observational data for multiple populations in GCs \citep[see e.g.,][]{libralato_2022,martens_2023,cordoni_2020,cordoni_2025}. For each model, and at each time, we have calculated the projected 2D radius ($R$) and the radial and tangential proper motions ($v_{\text{pmR}}$, $v_{\text{pmT}}$) for fifty different lines of sight. 

The analysis of these “observables” is not meant to be directly compared with observations but rather to illustrate how these measurements can trace the degree of dynamical mixing of the stellar populations and how they are linked to the phase space mixing discussed in the previous sections. Therefore, we have not included kinematic errors or completeness effects, and we assume that the two populations are clearly distinguishable with no 2P (1P) star misclassified as belonging to the 1P (2P) population.

\subsection{Spatial differences}
\label{sec:proj-aplus}

To quantify the differences in concentration and spatial distribution of both populations, we use the \aplus{} parameter \citep[][]{alessandrini_2016,lanzoni_2016}. The \aplus{} parameter measures the area difference between the cumulative stellar count of two samples of stellar tracers within a specific radial range [usually to measure the differences in the spatial distributions of populations with different masses, see e.g. \cite{alessandrini_2016,lanzoni_2016} for a study using \aplus{} to measure the segregation of blue-stragglers relatively to main-sequence and red giant stars, and \cite{weatherford_2018,della_croce_2024} for a study using \aplus{} to quantify the degree of mass segregation of more massive main sequence stars and the link between \aplus{} and the black hole content in GCs]. For our case, we follow a similar formulation for multiple populations as the one introduced by \cite{dalessandro_2018}:
\begin{equation}
    A^{+}(R_{\text{max}}) = \int_{0}^{1}{\phi_{1P}(R')-\phi_{2P}(R') dR'}\,,
\end{equation}
where $\phi(R')$ is the cumulative radial distribution, $R'=R/R_{\text{max}}$ and $R_{\text{max}}$ is the maximum distance within $A^{+}$ is calculated. In this work we adopted $R_{\text{max}}=2R_{\text{h}}$. We also normalise the cumulative radial profile so that $\phi(R_{\text{max}})=1$.

\begin{figure}
    \centering
    \includegraphics[width=\figwidth\linewidth]{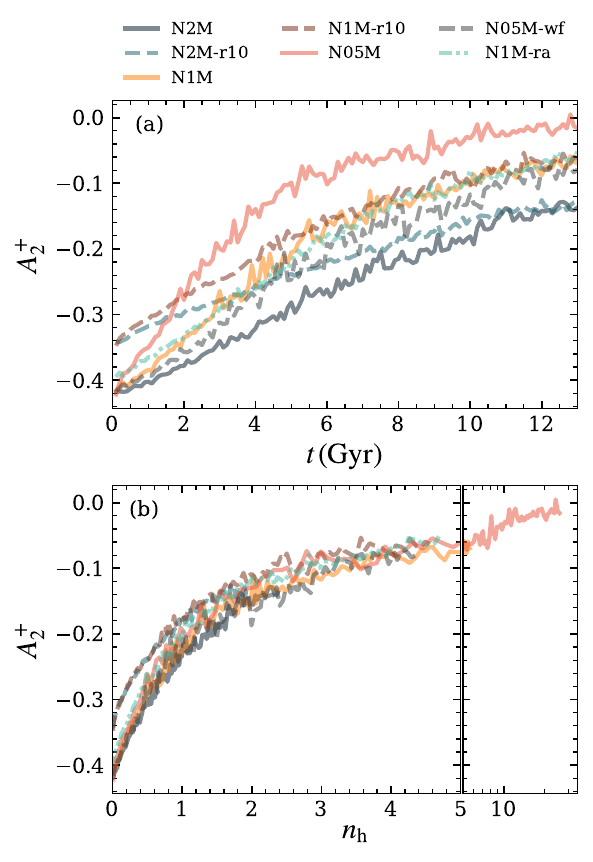}
    \caption{\aplus{2} parameter evolution for all GCs as a function of time (a) and number of relaxation times (b). All models follow a similar evolutionary path as the two populations spatially mix. The panel (a) scatter reduces when we race the cluster dynamical ages. After one relaxation time, the initial difference between the $r_h$ ratio of $20$ (N2M, N1M, N1M-ra, N05M and N05M-wf) and $r_h$ ratio of $10$ models (N2M-r10 and N1M-r10) is indistinguishable.}
    \label{fig:aplus2_all}
\end{figure}

Figure \ref{fig:aplus2_all} shows the evolution of the \aplus{2} parameter for all models as a function of time (panel (a)) and $n_h$ (panel (b)). If we look at a specific time, such as at \tms{12}{Gyr}, we find that our models have developed different degrees of spatial mixing: models N2M and N2M-r10 show the highest difference in concentration (more negative \aplus{2}), while model N05M has the most spatial mixing of the models (\aplus{2}$\sim 0$). On the other hand, if we look at the dynamical age $n_h$, we find that the picture changes significantly.  After $n_h\sim 1$, all models converge, and differences between the \aplus{2} parameter for the various models are less significant.

It is interesting to note that the initial concentration differences between the r10 models and some of the r20 models have mostly disappeared by a dynamical age $n_h \sim 1-2$; moreover, in all the models, the initial spatial differences between the 1P and the 2P populations significantly decrease during the clusters' early evolutionary phases; even in dynamically young clusters (for example in clusters with  $n_h \sim 1-3$), these spatial differences are significantly smaller than those in the initial conditions imprinted by the formation process. These results show that the strength of the present-day differences can not be used to infer the extent of the primordial differences and that even small present-day differences in dynamically young clusters may actually be the evolutionary outcome of multiple populations forming with strong dynamical differences [see also \cite{tiongco_2019,vesperini_2021,sollima_2021} and the discussion in \cite{caledano_2024}].

\begin{figure}
    \centering
    \includegraphics[width=\figwidth\linewidth]{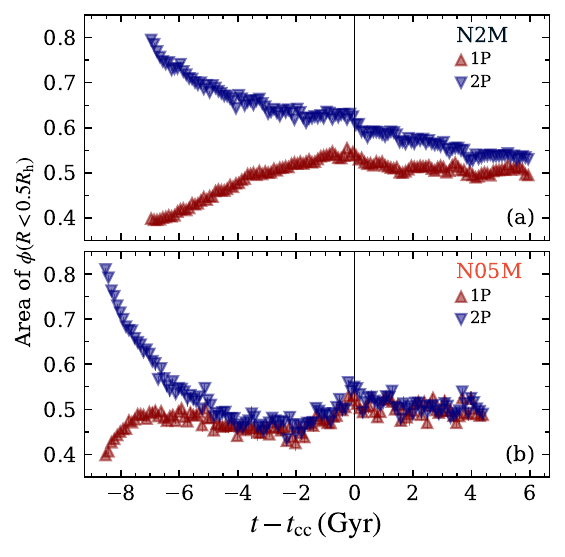}
    \caption{Area of the cumulative stellar distribution within $0.5R_{h}$ ($\phi(<05R_{h})$) for the two populations in models: (a) N2M and (b) N05M. As in Figure \ref{fig:s2-corecollapse}, we have centred the time evolution at the respective time of the first core collapse. The two populations in model N2M have not achieved complete spatial mixing and $\phi_{2P}(R<0.5R_{h})>\phi_{1P}(R<0.5R_{h})$. The model N05M has already attained a significant spatial mixing and $\phi_{2P}(R<0.5R_{h})\sim \phi_{1P}(R<0.5R_{h})$. Furthermore, for model N05M, both populations show the same rate of increased concentration at core collapse, which explains the lack of signatures in $S^2(E)$ (see Figure \ref{fig:s2-corecollapse}).}
    \label{fig:area_pop_cc}
\end{figure}

We do not find any clear signature of core collapse in the evolution of \aplus{} in Figure \ref{fig:aplus2_all}. However, we can see some interesting behaviour if we look at the individual areas of the cumulative stellar distribution, particularly in the cluster centre ($R<0.5R_{h}$). In Figure \ref{fig:area_pop_cc}, we show the area of the cumulative stellar distribution within $0.5R_{h}$ for the models N2M (panel a) and N05M (panel b), following the comparison in Figure \ref{fig:s2-corecollapse}. As we mentioned before in Section \ref{sec:internal_mixing_cc}, the two populations in the N05M have already spatially mixed before the cluster undergoes core collapse. At core collapse, the two populations increase their concentration at a similar rate. On the other hand, the model N2M has not spatially mixed by the time of the core collapse, and the second population has a higher concentration than the first population, as shown by the larger value for the area of $\phi(R<0.5R_{h})$.

\subsection{Velocity anisotropy differences}
\label{sec:proj-ani}

In this section, we will focus on the differences between the velocity anisotropy profiles of 1P and 2P stars. Several observational investigations of the velocity anisotropy profile of multiple populations have shown a 2P with varying degrees of radial anisotropy and a 1P that is isotropic or slightly tangential \citep[see][]{bellini_2015,cordoni_2020,cordoni_2025,libralato_2023,dalessandro_2024}. Multiple works aimed at studying the evolution of the radial profiles of the velocity dispersion and anisotropy have shown that the observed velocity anisotropy profiles are consistent with those found in systems that assume initial conditions characterized by a tidally filling 1P system and a tidally underfilling 2P subsystem [see, e.g., \cite{vesperini_2021,sollima_2021,livernois_2024}; see also e.g. \cite{tiongco_2016} and references therein for a similar evolution of the anisotropy in tidally filling and tidally underfilling single-population clusters].

Our goal is to quantify these differences; therefore, we adopt a new parameter that provides a global measure of the difference between the velocity anisotropy of the two populations by means of the area between the best-fitted velocity anisotropy models for the 1P and the 2P. We first calculate the velocity dispersion profiles for the radial and tangential components (projected on the sky) from the selected sample of stars. Then, we fit a parametric model for the radial velocity dispersion and the velocity anisotropy. For the radial velocity dispersion, we use a 4th-order polynomial following the parameterisation in \cite{watkins_2022}, then for the tangential component, we define:
\begin{equation}
\sigma_{\text{tan}}^2 = \sigma_{\text{rad}}^2(1-\beta_{\text{2D}})\,,    
\end{equation}
where we define the projected velocity anisotropy profile, $\beta_{\text{2D}} = 1- \sigma_{\text{tan}}^2/\sigma_{\text{rad}}^2$, with the parameterisation: 
\begin{equation}
\beta_{\text{2D}}(R) =      
\begin{cases}
\frac{\beta_{i}R^2}{(R_{\text{a}}^2+R^2)}\left(1-\frac{R}{R_{\text{t}}}\right) & \text{for}\, R < R_{\text{t}}\\
0 &\text{for}\, R \geq R_{\text{t}}\,.
\end{cases}
\end{equation}
This parameterization is based on the Osipkov-Merrit velocity anisotropy profile \citep{osipkov_1979,merrit_1985} with the anisotropy radius $R_{\text{a}}$, but allowing for tangential anisotropy ($\beta_{i}<0$) and a maximum radial anisotropy smaller than $\max(\beta_{\text{2D}}(R))<1$. We have also included a truncation term, $(1-R/R_{\text{t}})$, so that the anisotropy profile returns to being isotropic ($\beta_{\text{2D}}=0$) at the truncation radius $R_{\text{t}}$. The truncation term allows us to properly characterise our simulations' velocity anisotropy profile at large radii. For the purpose of this analysis, we have fixed the anisotropy truncation radius at the respective tidal radius of each simulated cluster’s snapshot. We then calculate the anisotropy area $A_{\beta}$ for each population by first normalising the radius with the tidal radius and then integrating over $\beta_{\text{2D}}(R)$.

\begin{figure}
    \centering
    \includegraphics[width=\figwidth\linewidth]{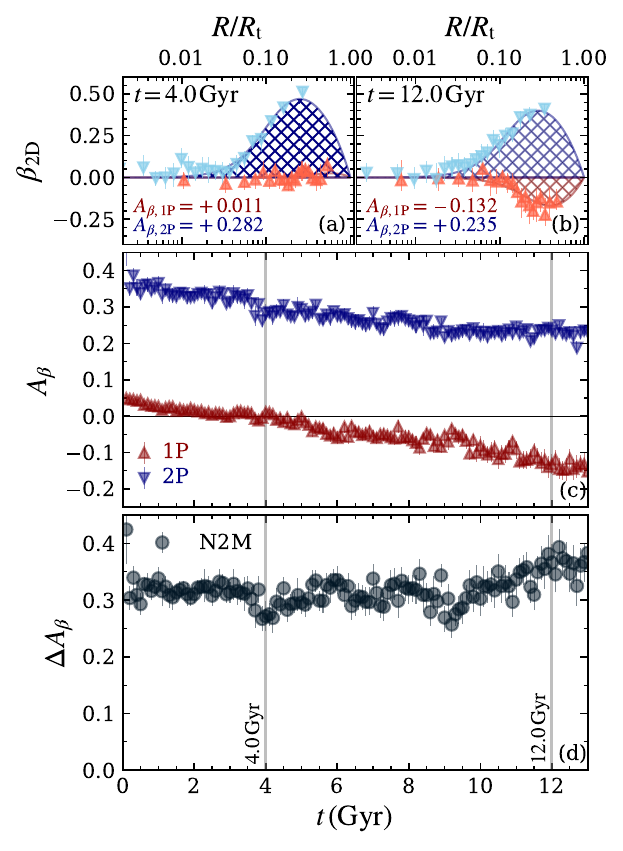}
    \caption{Velocity anisotropy area ($A_{\beta}$) and area difference ($\Delta A_{\beta}$) for the N2M model. Panels (a) and (b) show the velocity anisotropy for the two populations (red upper triangles, PO1P, and downwards blue triangles, PO2P) in the N2M model at \tms{4}{Gyr} and \tms{12}{Gyr}, respectively. The hatched regions represent the respective anisotropy area for each population from the best-fit velocity dispersion and anisotropy parameterisation (see main text). Panel (c) shows the time evolution of the anisotropy areas for the two populations over time; as the second population becomes less radially anisotropic, the first population becomes tangentially anisotropic and keeps a constant difference with the second population, as shown in panel (d), where we show the evolution of the anisotropy area difference. The vertical lines in panels (c) and (d) represent the times for which we show the velocity anisotropy profiles in panels (a) and (b).}
    \label{fig:ani_areas_example}
\end{figure}

Panels (a) and (b) in Figure \ref{fig:ani_areas_example} show the best-fit velocity anisotropy profiles for the N2M model at two different times (\tml{4}{Gyr} and \tml{12}{Gyr}, respectively). At \tms{4}{Gyr}, the second population shows radial anisotropy with an anisotropy area of $A_{\beta,\text{2P}}=0.282$ while the first population is close to be isotropic with $A_{\beta,\text{1P}}=0.011$. At \tms{12}{Gyr} the second population has become slightly less radially anisotropic with $A_{\beta,\text{2P}}=0.235$, and the first population is now tangentially anisotropic with $A_{\beta,\text{1P}}=-0.132$. In panel (c), we show the time evolution of the anisotropy areas for model N2M. Both populations started isotropic and had $A_{\beta}=0$; during the first \tms{100}{Myr} of evolution, both populations developed radial anisotropy (with a stronger anisotropy for the 2P population). The development of radial anisotropy follows the early mass loss due to stellar evolution, followed by a shallowing of the potential and an expansion of the cluster. The difference in the strength of the radial anisotropy is a consequence of the initial extension of the populations and the tidally filling and underfilling conditions for the first and second populations, respectively. As the cluster evolves, the second population steadily loses the radial anisotropy, while the first population becomes isotropic and then develops tangential anisotropy.

Panel (d) in Figure \ref{fig:ani_areas_example} shows the anisotropy area difference between the populations for the N2M model. This difference is defined as:
\begin{equation}
    \Delta A_{\beta} = A_{\beta,\text{2P}} - A_{\beta,\text{1P}}\,,
\end{equation}
if both populations are close to isotropic ($A_{\beta,\text{2P}}\sim A_{\beta,\text{1P}}\sim 0$) or have similar anisotropy profiles ($A_{\beta,\text{2P}} \sim A_{\beta,\text{1P}}\neq 0$), then $\Delta A_{\beta}$ is close to zero, and we will consider the populations to be kinematically mixed. On the other hand, as $\Delta A_{\beta}$ becomes different from zero, the populations are further from mixing. This could be the case when 2P has a strong radial anisotropy, and 1P is isotropic (panel (a) in Figure \ref{fig:ani_areas_example}), or the two populations have opposite velocity anisotropy (as in panel (b) in Figure \ref{fig:ani_areas_example} where the 2P is radially anisotropic and the 1P tangentially anisotropic). We see that for the N2M model, the anisotropy area difference stays relatively constant, consistent with it being the model further from mixing at the end of the simulations. 

\begin{figure}
    \centering
    \includegraphics[width=\figwidth\linewidth]{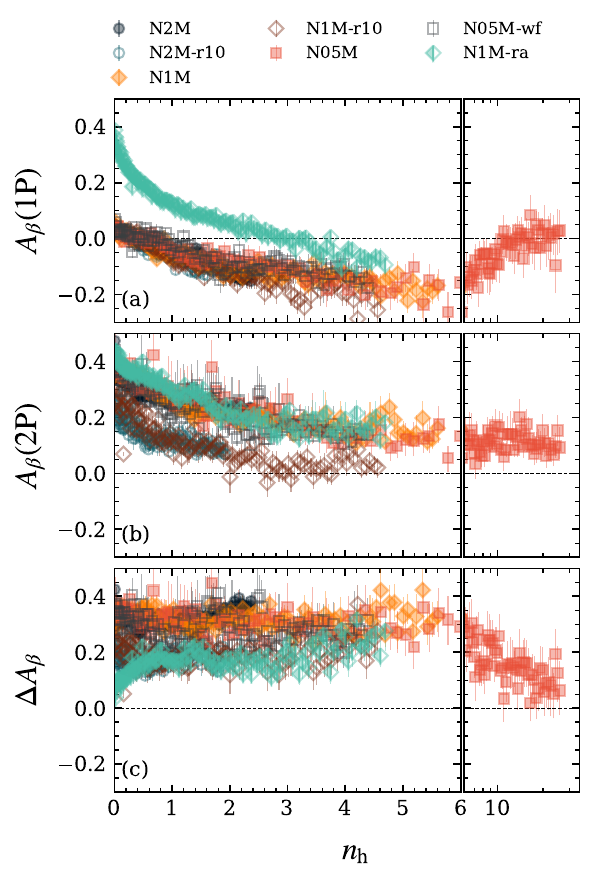}
    \caption{Anisotropy area ($A_{\beta}$) and area difference ($\Delta A_{\beta}$) as described in Figure \ref{fig:ani_areas_example}, for all models in our sample with respect to their dynamical ages. For a better visualization, we separated $n_{\text{h}}$ in a linear section ($n_{\text{h}}\leq6$) and a logarithmic section ($n_{\text{h}}>6$).}
    \label{fig:ani_areas_all}
\end{figure}

We show the evolution of the anisotropy area for each population and the anisotropy difference for all models in Figure \ref{fig:ani_areas_all}. Our choice of initial conditions affects the populations differently. In panel (a), the first population for all models starting isotropically follow the same path: starting at $A_{\beta}=0$, developing a very mild radial anisotropy, becoming isotropic and then developing a tangential anisotropy. Only the N05M model returns to being close to isotropic at the end of the simulation. On the other hand, the N1M-ra model's first population starts with a primordial radial anisotropy, which translates to an initial $A_{\beta} = 0.4$, the cluster’s first population proceeds to lose the radial anisotropy and becomes isotropic at \nh{3}. The evolution of the velocity anisotropy in the first population is driven by its tidally filling condition and is consistent with tidally filling single population simulations \citep[see e.g.,][]{tiongco_2016}. In panel (b), the second population shows two main paths of evolution driven by its initial concentration: as the second population expands, clusters with an initially more concentrated second population develop a more substantial radial anisotropy ($A_{\beta}\sim0.4$) than those with an initial concentration of r10 ($A_{\beta}\sim 0.3$). As for the N1M-ra model, note that, given the primordial velocity anisotropy profile of this model, the 2P stars populate mainly the inner isotropic regions, while the radial anisotropy affects mainly the 1P populations extending in the outer anisotropic regions.

In panel (c) of Figure \ref{fig:ani_areas_all}, we show the area difference for each model. Following the discussion for panels (a) and (b), the imprint of the initial conditions is still present on the area difference. The initially isotropic models with r20 concentration develop a $\Delta A_{\beta}\sim0.35$ after the first \tms{100}{Myr}, while for the models with r10, $\Delta A_{\beta}\sim0.25$. We see that for the two concentration groups, $\Delta A_{\beta}$ slowly decreases as the clusters become dynamically older, showing that differences in velocity anisotropy persist over several relaxation times. Only after \nh{6} does the N05M model show a continuous decline in $\Delta A_{\beta}$, getting close to complete kinematic mixing. On the other hand, the N1M-ra model has a $\Delta A_{\beta}=0.05$ after the first \tms{100}{Myr} starting in a “close to kinematic mixing” state, this is because the second population develops the radial anisotropy while the first population was already radially anisotropic. As the cluster evolves dynamically and the first population becomes isotropic, the $\Delta A_{\beta}$ value of the N1M-ra model increases and becomes similar to those with an initial concentration of r10. By the end of the simulation, as the first population of the N1M-ra model has become tangentially anisotropic, its $\Delta A_{\beta}$ value is now consistent with the r20 models. 

In \cite{dalessandro_2024}, we estimated the anisotropy area (referred to as $\alpha_{\beta}$ in that paper) and anisotropy area difference (referred to as $(\alpha_{\beta}^{SP}-\alpha_{\beta}^{FP})$ in that paper) for a sample of Galactic globular clusters from their observed kinematics, using the same definitions described here, and we showed the N1M, N1M-ra and N05M models for a comparison between simulations and observations. Interestingly, we find a consistent trend between the observations and simulations in terms of the evolution of these quantities. While further comparisons of the observed GCs and the simulations are beyond the scope of our current sample of simulations, we stress here the usefulness of the anisotropy area difference to trace the mixing process.

\subsection{Projected angular momentum differences}
\label{sec:proj-Lz}

In Section \ref{sec:internal_mixing}, we showed that differences in the marginalised angular momentum can also trace the kinematic mixing (see Figures \ref{fig:phase-space-mixing-example}, \ref{fig:s2-corecollapse}, \ref{fig:s2-anisotropy} and \ref{fig-app:all_diff_3d}). Furthermore, differences in the velocity anisotropy imply differences in the radial and tangential velocity dispersion components that can manifest themselves in the angular momentum distribution. As initially hinted in Section \ref{sec:internal_mixing_ani}, stars on more radial orbits will have smaller angular momentum at a given radius. 

In this section, we explore the connection between the projected angular momentum  $L_{\text{z}} =v_{T}\times R$ (where $v_{T}$ is the tangential velocity on the plane of the sky, and $R$ is the projected distance from the cluster's centre), and the velocity anisotropy. In a non-rotating cluster, the $L_{\text{z}}$ distribution at any given radius is centred at zero, and, therefore, we focus on the dispersion of the $L_{\text{z}}$ distribution, $\sigma_{L_{\text{z}}}$, as an observational tracer of the tri-dimensional angular momentum. In the same way as for the angular momentum in Section \ref{sec:internal_mixing}, we normalise by the half-mass angular momentum $L_{\text{h}}$ of a circular orbit around an equivalent point-mass potential. 

\begin{figure}
    \centering
    \includegraphics[width=\figwidth\linewidth]{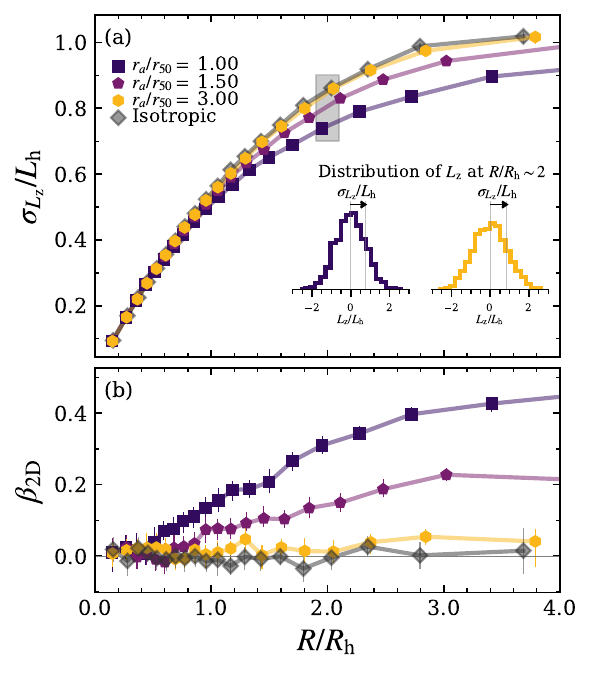}
    \caption{Projected angular momentum dispersion $\sigma_{L_{\text{z}}}$ profiles for King models with different velocity anisotropy profiles (panel (a)) and their respective velocity anisotropy profiles (panel (b)). As the King models become more radially anisotropic (smaller values of $r_{a}/r_{h}$), the projected angular momentum dispersion decreases. The insets on panel (a) show the distribution of the projected angular momentum at $r\sim2R_{\text{h}}$ for the models with the most and least radial anisotropy. The grey area in panel (a) shows the range of radial distances of stars used to calculate the histograms in the panel. We have normalised $\sigma_{L_{\text{z}}}$ by the half-mass angular momentum $L_{\text{h}}$, corresponding to the angular momentum of a circular orbit of an equivalent point-mass potential (see Section \ref{sec:internal_mixing}).}
    \label{fig:sigma_Lz_king}
\end{figure}

In order to provide some initial guidance on the link between $\sigma_{L_{\text{z}}}$ and the cluster's velocity anisotropy, we show in Figure \ref{fig:sigma_Lz_king} the radial profile of $\sigma_{L_{\text{z}}}$ for four single-population King models with different radial anisotropy profiles built with the \texttt{Limepy} code \citep{gieles_2015,gieles_2017}. The value of $\sigma_{L_{\text{z}}}$ at a given radius in the cluster's outer regions decreases for more radially anisotropic models. This decrease in $\sigma_{L_{\text{z}}}$ for more radially anisotropic velocity distributions is consistent with the expected smaller fraction of orbits with larger angular momentum (corresponding to a narrower distribution of $L_\text{z}$ for more anisotropic models).

\begin{figure}
    \centering
    \includegraphics[width=\figwidth\linewidth]{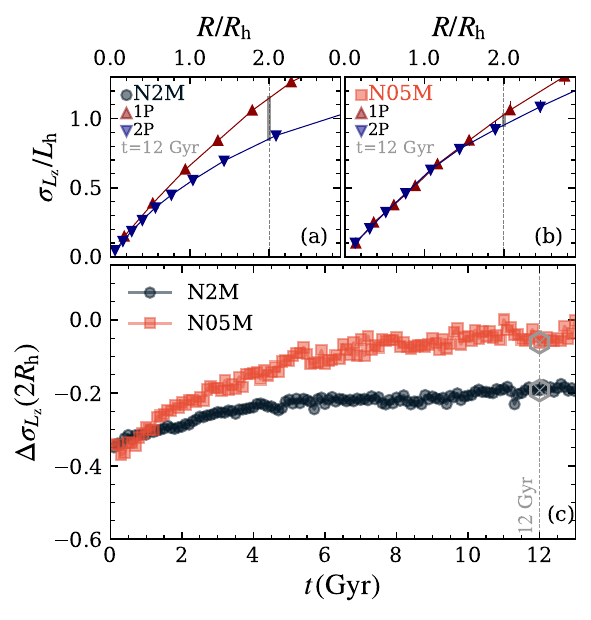}
    \caption{Projected angular momentum dispersion $\sigma_{L_{\text{z}}}$ profiles for both populations in the N2M model (panel (a)) and the N05M model (panel(b)) at \tms{12}{Gyr}. The vertical grey solid lines in panels (a) and (b) represent the difference in $\sigma_{L_{\text{z}}}$ between the populations at $R=2R_{\text{h}}$. Panel (c) shows the time evolution of $\sigma_{L_{\text{z}}}$ difference at $2R_{\text{h}}$; the closer this value is to zero, the closer the two populations are to be kinematically mixed.}
    \label{fig:sigma_Lz_example}
\end{figure}

Following the correlations observed in the single population King models, the differences in velocity anisotropy between the stellar populations discussed in Section \ref{sec:proj-ani} should also appear as differences in $\sigma_{L_{\text{z}}}$.  Figure \ref{fig:sigma_Lz_example} shows the $\sigma_{L_{\text{z}}}$ profile for the two populations of the N2M and N05M models at \tms{12}{Gyr} (panels (a) and (b)). The N2M model, which also shows the largest velocity anisotropy area difference, also shows that the $\sigma_{L_{\text{z}}}$ profiles are different and as expected, $\sigma_{L_{\text{z}}}(1P) \geq \sigma_{L_{\text{z}}}(2P)$. The N05M model shows almost no differences between the $\sigma_{L_{\text{z}}}$ profiles, consistent with being close to complete mixing. Considering how the differences in $\sigma_{L_{\text{z}}}$ appear more significant outside of the half-light radius, we follow the time evolution of the difference between the 1P and 2P values of $\sigma_{L_{\text{z}}}$ at $r=2R_{\text{h}}$ for all epochs in our simulations to follow its evolution. Panel (c) shows the time evolution of $\Delta\sigma_{L_{\text{z}}}$ for models N2M and N05M. This plot shows that, as the cluster evolves and mixes, $\Delta\sigma_{L_{\text{z}}}$ approaches zero and provides another tracer to measure the kinematic mixing of the two populations.

\begin{figure}
    \centering
    \includegraphics[width=\figwidth\linewidth]{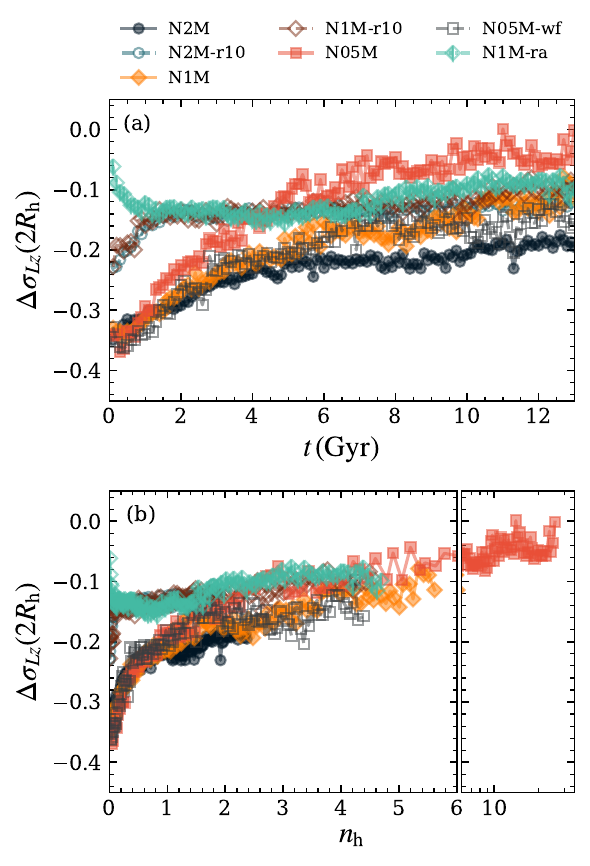}
    \caption{Evolution of the projected angular momentum difference at $2R_{\text{h}}$ for all models in our sample as described in panel (c) of Figure \ref{fig:sigma_Lz_example}. In panel (a) with respect to the physical time of the simulation and in panel (b) with respect to their dynamical age. Once we consider the dynamical age, all models converge into the same evolutionary path, independently of their initial conditions. }
    \label{fig:sigma_Lz_all}
\end{figure}

Figure \ref{fig:sigma_Lz_all} shows the evolution of $\Delta\sigma_{L_{\text{z}}}$ for all models in our sample in terms of the physical time and dynamical age. The models N2M and N05M show the most and least differences in $\sigma_{L_{\text{z}}}$ between the populations; the models with initial r10 concentration start with smaller values of $\Delta\sigma_{L_{\text{z}}}$ than the r20 models, and after \tms{1}{Gyr} they overlap with the N1M-ra model. The latter also moves to a larger difference in $\sigma_{L_{\text{z}}}$, consistent with the first population evolving from radially anisotropic to isotropic and then tangentially anisotropic. If we look now at the dynamical age, the evolutionary paths change, and instead of having a set of parallel tracks, all models tend to converge into a single path. The models N1M-ra, N1M-r10, and N2M-r10 join in a single path at about \nh{0.1}, early in the dynamical evolution of the clusters. As we are measuring the differences at $2R_{\text{h}}$, we have a degeneration between the effects of a weaker radially anisotropic second population (N1M-r10 and N2M-r10) and two populations with different degrees of radial anisotropy (N1M-ra, the first population becomes isotropic at \nh{3}, see panel(a) in Figure \ref{fig:ani_areas_all}). The rest of the models follow a single path and start to overlap at \nh{4}. While most of our models do not get past this point, we expect they will follow the path of the N05M model.

\subsection{Connection with the $S^2$ parameter}
\label{sec:proj-s2}

We have explored three different observational parameters to quantify the degree of mixing of the multiple populations. All three of them trace different aspects of the mixing process: \aplus{2} follows the spatial mixing, while $\Delta A_{\beta}$ and $\Delta\sigma_{L_{\text{z}}}$ trace differences in the velocity distribution anisotropy of the two populations and follow their kinematic mixing. In section \ref{sec:internal_mixing}, we defined the $S^2$ parameter as a way to describe the overall phase-space mixing. Now, we compare the $S^2$ parameter, which cannot be observed directly, with the three observable tracers of mixing.

\begin{figure}
    \centering
    \includegraphics[width=\figwidth\linewidth]{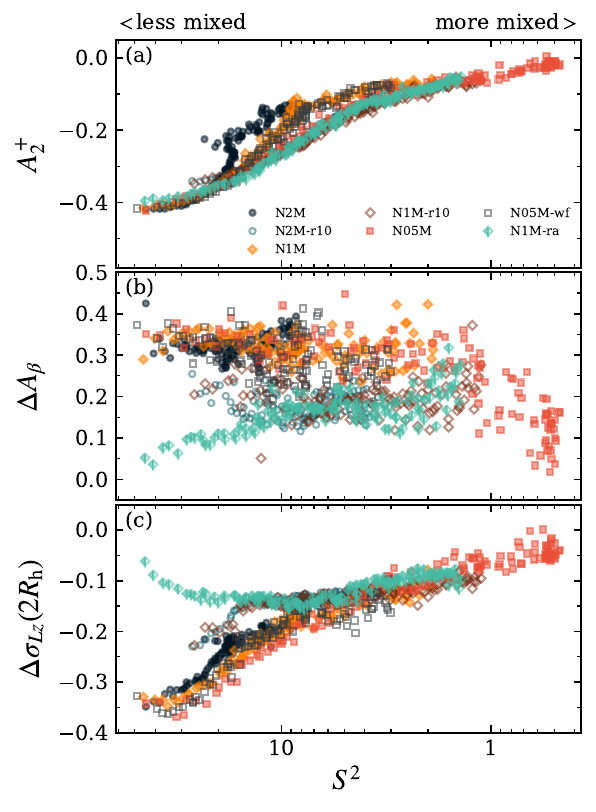}
    \caption{Comparison of the three projected parameters that quantify the degree of mixing of the multiple populations and the $S^2$ parameter from the phase-space mixing. The $S^2$ parameter decreases to the right following the increasing phase-space mixing of the two populations, as stated at the top of the panel (a).}
    \label{fig:s2_obs}
\end{figure}

Figure \ref{fig:s2_obs} shows the three projected quantities we analysed as a function of the $S^2$ parameter, with increasing mixing, i.e. decreasing $S^2$ towards the right. We find the exact overall behaviour when comparing each parameter with the dynamical time. As the clusters become more mixed (smaller value of $S^2$), the observational parameters approach zero. However, we can also distinguish additional features due to the cluster's different evolutionary state. Panel (a) of Figure \ref{fig:s2_obs} shows the effects of core collapse and the subsequent separation in different evolutionary paths due to the degree of spatial mixing in the core at core collapse and the path for the clusters that do not undergo core collapse. The latter do not show the temporary increase of $S^2$ (corresponding to the bump shown in Figure \ref{fig:s2-comparison}). In the case of the velocity anisotropy area difference in panel (b), we see similar behaviour to that in Figure \ref{fig:ani_areas_all}. Finally, for the difference in the projected angular momentum in panel (c), all models start converging to the same path after attaining a level of mixing of $S^2\simeq9$. It is important to note that at this value of $S^2\simeq9$, there are still differences between models when looking at the velocity anisotropy area difference.

\section{Discussion and summary}
\label{sec:summary}

In this work, we analysed the long-term evolution of simulations with multiple populations to characterise the dynamical mixing process. Our simulations follow the evolution of models with an initially concentrated and underfilling second population in an extended, more massive and tidally filling first population. We explore different initial conditions to examine differences in the concentration of the second population, primordial velocity anisotropy of the cluster and the strength of the tidal field. 

Multiple studies have shown that the different stellar populations in GCs have a large variety of spatial and kinematic mixing as traced by differences in concentration \citep[see e.g.,][]{dalessandro_2019,leitinger_2023,caledano_2024}, rotation \citep[see e.g.,][]{cordero_2017,cordoni_2020,dalessandro_2021,dalessandro_2024,martens_2023,leitinger_2025} and velocity anisotropy \citep[see e.g.,][]{richer_2013,bellini_2015,cordoni_2020,cordoni_2025,libralato_2023,dalessandro_2024}. To trace the evolution of the mixing processes, we followed the differences in the phase-space distribution of each population in our models. We defined the $S^2$ parameter as the sum of the energy and angular momentum distribution square differences over the whole phase space. The $S^2$ parameter allows us to build a comprehensive dynamical picture of the mixing process, including both the spatial and kinematic components of the populations' mixing. We find that (as expected) mixing increases with dynamical age (see Figure \ref{fig:s2-comparison}).  We also show that core-collapse and primordial velocity anisotropy can alter the mixing rate of the cluster and core collapse may even cause a temporary slight inversion of the phase-space mixing (see Figures \ref{fig:s2-corecollapse} and \ref{fig:s2-anisotropy}). As shown in Figure \ref{fig:s2-comparison}, models can achieve similar values of $S^2$ at different physical and dynamical ages, either due to variations in their mixing rates or initial conditions. Our models reach different degrees of dynamical mixing at $12\,\text{Gyr}$; among all the models presented, the N05M and N2M exhibit, respectively, the most and least mixing.

We followed the evolution of three “observable” parameters that can trace the degree of spatial and kinematic mixing and found they follow the same general trends found in previous observational studies \citep[see, e.g.][]{dalessandro_2019,dalessandro_2024,caledano_2024}, where the differences in the spatial and kinematic properties of multiple populations decrease with dynamical age. We also find that in our models, the strength of the initial 1P and 2P differences decreases already in the clusters' early evolutionary phases when the clusters are still dynamically young ($n_h \sim 1-2$) \citep[see also][]{vesperini_2021,sollima_2021,caledano_2024}.
Present-day differences still contain important information about the 1P and 2P dynamical properties imprinted at the time of the cluster formation. However, it is important to note that the observed present-day strength of these differences can be much smaller than the initial ones, even in dynamically young clusters, and that different initial conditions may lead to similar present-day properties. Furthermore, and as a cautionary note, our data do not consider limitations typically affecting observational analysis, such as incompleteness or uncertainties in the population membership, which can add stochasticity to the evolutionary paths and reduce the detected spatial differences shown in Figure \ref{fig:aplus2_all}.

We quantify the difference in the velocity anisotropy profiles of each population with the $\Delta A_{\beta}$ parameter, which measures the area difference between the velocity anisotropy profiles. This parameter is the most lasting kinematic signature for mixing, as the models still keep differences after multiple relaxation times. The overall trend is driven by a radially anisotropic second population and a tangentially anisotropic first population. For all the initially isotropic models, the 1P remains isotropic in the inner regions and develops tangential anisotropy in the outer regions, while 2P evolves towards a radially anisotropic velocity distribution with the strength of the radial anisotropy depending on the 2P initial concentration: initially less concentrated 2P spatial distributions results into a shallower radial velocity anisotropy (see Figure \ref{fig:ani_areas_all}). The model with an initial radially anisotropic velocity distribution illustrates the complexity of the interplay between the effects due to different spatial and velocity distributions. In the initially anisotropic model we have studied (N1M-ra with $r_{\text{h},1\text{P}}/r_{\text{h}2,\text{P}}=20$) the difference between the 1P and 2P anisotropy, after $n_h \sim 1$ evolves similarly to that of an initially isotropic model but with smaller initial spatial differences (N1M-r10). Additional models are necessary to further explore these aspects of the clusters' dynamical mixing. 

Finally, we have studied the evolution of $\sigma_{L_{\text{z}}}$, the dispersion of the angular momentum calculated from the tangential component of the proper motion velocity on the plane of the sky and the projected radial distance from the cluster's centre. We have shown that this parameter is linked to the degree of radial anisotropy and the differences between the values of this parameter for the 1P and the 2P populations, $\Delta\sigma_{L_{\text{z}}}$, trace the differences in the degree of the anisotropy of their velocity distributions; but more importantly, this parameter allows for a more direct way to quantify differences in the velocity distribution of the two populations and provides another observational measure of the kinematics of multiple populations and the degree of kinematic mixing. We have explored its evolutionary path (see Figures \ref{fig:sigma_Lz_example} and \ref{fig:sigma_Lz_all}) as well as its co-evolution with the global phase space mixing measured by $S^2$ (see Figure \ref{fig:s2_obs}). For dynamical ages $n_h \gtrsim 1-2$, the evolutionary paths of $\Delta\sigma_{L_{\text{z}}}$ for models with different initial conditions approach a common universal track. As pointed out above, dynamical differences observed in old clusters provide key information on the differences between the 1P and the 2P spatial and kinematic properties set by the formation processes and their evolution, but, even for dynamically young clusters, it is necessary to exercise caution in interpreting the specific strength of the present-day differences as close indication of those imprinted at the time of the multiple population formation.

Models extending the investigation presented here to a broader set of initial conditions are necessary to further explore the evolutionary path of the parameters introduced in this paper and their link with the clusters' initial conditions. Including other dynamical indicators, such as the differences in the degree of energy equipartition and internal rotation, will provide additional key steps towards a comprehensive picture of the dynamical mixing of multiple populations in globular clusters.

\begin{acknowledgements}
EV and FA acknowledge support from STScI grant AR-16157. EV acknowledges also support from the John and A-Lan Reynolds Faculty Research Fund. ED acknowledges financial support from the Fulbright Visiting Scholar program 2023. ED is also grateful for the warm hospitality of Indiana University, where part of this work was performed. This research was supported in part by Lilly Endowment, Inc., through its support for the Indiana University Pervasive Technology Institute. FA also acknowledges the following Python packages used during the research presented here: \texttt{NumPy} \citep{numpy_2020}, \texttt{SciPy} \citep{scipy_2020}, \texttt{emcee} \citep{emcee_2013} and \texttt{Astropy} \citep{astropy_2022}, while all figures were made using \texttt{Matplotlib} \citep{matplotlib_2007}.   
\end{acknowledgements}

\bibliographystyle{aa} 
\bibliography{multipop_arxiv_ver} 

\begin{thebibliography}{70}
\expandafter\ifx\csname natexlab\endcsname\relax\def\natexlab#1{#1}\fi

\bibitem[{{Alessandrini} {et~al.}(2016){Alessandrini}, {Lanzoni}, {Ferraro},
  {Miocchi}, \& {Vesperini}}]{alessandrini_2016}
{Alessandrini}, E., {Lanzoni}, B., {Ferraro}, F.~R., {Miocchi}, P., \&
  {Vesperini}, E. 2016, \apj, 833, 252

\bibitem[{{Amaro-Seoane} {et~al.}(2013){Amaro-Seoane}, {Konstantinidis},
  {Brem}, \& {Catelan}}]{amaro_seoane_2013}
{Amaro-Seoane}, P., {Konstantinidis}, S., {Brem}, P., \& {Catelan}, M. 2013,
  \mnras, 435, 809

\bibitem[{{Astropy Collaboration} {et~al.}(2022){Astropy Collaboration},
  {Price-Whelan}, {Lim}, {Earl}, {Starkman}, {Bradley}, {Shupe}, {Patil},
  {Corrales}, {Brasseur}, {N{\"o}the}, {Donath}, {Tollerud}, {Morris},
  {Ginsburg}, {Vaher}, {Weaver}, {Tocknell}, {Jamieson}, {van Kerkwijk},
  {Robitaille}, {Merry}, {Bachetti}, {G{\"u}nther}, {Aldcroft},
  {Alvarado-Montes}, {Archibald}, {B{\'o}di}, {Bapat}, {Barentsen},
  {Baz{\'a}n}, {Biswas}, {Boquien}, {Burke}, {Cara}, {Cara}, {Conroy},
  {Conseil}, {Craig}, {Cross}, {Cruz}, {D'Eugenio}, {Dencheva}, {Devillepoix},
  {Dietrich}, {Eigenbrot}, {Erben}, {Ferreira}, {Foreman-Mackey}, {Fox},
  {Freij}, {Garg}, {Geda}, {Glattly}, {Gondhalekar}, {Gordon}, {Grant},
  {Greenfield}, {Groener}, {Guest}, {Gurovich}, {Handberg}, {Hart},
  {Hatfield-Dodds}, {Homeier}, {Hosseinzadeh}, {Jenness}, {Jones}, {Joseph},
  {Kalmbach}, {Karamehmetoglu}, {Ka{\l}uszy{\'n}ski}, {Kelley}, {Kern},
  {Kerzendorf}, {Koch}, {Kulumani}, {Lee}, {Ly}, {Ma}, {MacBride}, {Maljaars},
  {Muna}, {Murphy}, {Norman}, {O'Steen}, {Oman}, {Pacifici}, {Pascual},
  {Pascual-Granado}, {Patil}, {Perren}, {Pickering}, {Rastogi}, {Roulston},
  {Ryan}, {Rykoff}, {Sabater}, {Sakurikar}, {Salgado}, {Sanghi}, {Saunders},
  {Savchenko}, {Schwardt}, {Seifert-Eckert}, {Shih}, {Jain}, {Shukla}, {Sick},
  {Simpson}, {Singanamalla}, {Singer}, {Singhal}, {Sinha}, {Sip{\H{o}}cz},
  {Spitler}, {Stansby}, {Streicher}, {{\v{S}}umak}, {Swinbank}, {Taranu},
  {Tewary}, {Tremblay}, {de Val-Borro}, {Van Kooten}, {Vasovi{\'c}}, {Verma},
  {de Miranda Cardoso}, {Williams}, {Wilson}, {Winkel}, {Wood-Vasey}, {Xue},
  {Yoachim}, {Zhang}, {Zonca}, \& {Astropy Project
  Contributors}}]{astropy_2022}
{Astropy Collaboration}, {Price-Whelan}, A.~M., {Lim}, P.~L., {et~al.} 2022,
  \apj, 935, 167

\bibitem[{{Bastian} \& {Lardo}(2018)}]{bastian_2018}
{Bastian}, N. \& {Lardo}, C. 2018, \araa, 56, 83

\bibitem[{{Bekki}(2010)}]{bekki_2010}
{Bekki}, K. 2010, \apjl, 724, L99

\bibitem[{{Bekki}(2011)}]{bekki_2011}
{Bekki}, K. 2011, \mnras, 412, 2241

\bibitem[{{Bellini} {et~al.}(2015){Bellini}, {Vesperini}, {Piotto}, {Milone},
  {Hong}, {Anderson}, {van der Marel}, {Bedin}, {Cassisi}, {D'Antona},
  {Marino}, \& {Renzini}}]{bellini_2015}
{Bellini}, A., {Vesperini}, E., {Piotto}, G., {et~al.} 2015, \apjl, 810, L13

\bibitem[{{Berczik} {et~al.}(2025){Berczik}, {Panamarev}, {Ishchenko}, \&
  {Kocsis}}]{berczik_2025}
{Berczik}, P., {Panamarev}, T., {Ishchenko}, M., \& {Kocsis}, B. 2025, \aap,
  694, A163

\bibitem[{{Breen} {et~al.}(2017){Breen}, {Varri}, \& {Heggie}}]{breen_2017}
{Breen}, P.~G., {Varri}, A.~L., \& {Heggie}, D.~C. 2017, \mnras, 471, 2778

\bibitem[{{Cadelano} {et~al.}(2024){Cadelano}, {Dalessandro}, \&
  {Vesperini}}]{caledano_2024}
{Cadelano}, M., {Dalessandro}, E., \& {Vesperini}, E. 2024, \aap, 685, A158

\bibitem[{{Cai} {et~al.}(2016){Cai}, {Gieles}, {Heggie}, \& {Varri}}]{cai_2016}
{Cai}, M.~X., {Gieles}, M., {Heggie}, D.~C., \& {Varri}, A.~L. 2016, \mnras,
  455, 596

\bibitem[{{Calura} {et~al.}(2019){Calura}, {D'Ercole}, {Vesperini}, {Vanzella},
  \& {Sollima}}]{calura_2019}
{Calura}, F., {D'Ercole}, A., {Vesperini}, E., {Vanzella}, E., \& {Sollima}, A.
  2019, \mnras, 489, 3269

\bibitem[{{Cordero} {et~al.}(2017){Cordero}, {H{\'e}nault-Brunet},
  {Pilachowski}, {Balbinot}, {Johnson}, \& {Varri}}]{cordero_2017}
{Cordero}, M.~J., {H{\'e}nault-Brunet}, V., {Pilachowski}, C.~A., {et~al.}
  2017, \mnras, 465, 3515

\bibitem[{{Cordoni} {et~al.}(2025){Cordoni}, {Casagrande}, {Milone},
  {Dondoglio}, {Mastrobuono-Battisti}, {Jang}, {Marino}, {Lagioia}, {Legnardi},
  {Ziliotto}, {Muratore}, {Mehta}, {Lacchin}, {Tailo}, {Research School of
  Astronomy}, \& {Astrophysics}}]{cordoni_2025}
{Cordoni}, G., {Casagrande}, L., {Milone}, A.~P., {et~al.} 2025, \mnras, 537,
  2342

\bibitem[{{Cordoni} {et~al.}(2020){Cordoni}, {Milone}, {Mastrobuono-Battisti},
  {Marino}, {Lagioia}, {Tailo}, {Baumgardt}, \& {Hilker}}]{cordoni_2020}
{Cordoni}, G., {Milone}, A.~P., {Mastrobuono-Battisti}, A., {et~al.} 2020,
  \apj, 889, 18

\bibitem[{{Dalessandro} {et~al.}(2024){Dalessandro}, {Cadelano}, {Della Croce},
  {Aros}, {White}, {Vesperini}, {Fanelli}, {Ferraro}, {Lanzoni}, {Leanza}, \&
  {Origlia}}]{dalessandro_2024}
{Dalessandro}, E., {Cadelano}, M., {Della Croce}, A., {et~al.} 2024, \aap, 691,
  A94

\bibitem[{{Dalessandro} {et~al.}(2019){Dalessandro}, {Cadelano}, {Vesperini},
  {Martocchia}, {Ferraro}, {Lanzoni}, {Bastian}, {Hong}, \&
  {Sanna}}]{dalessandro_2019}
{Dalessandro}, E., {Cadelano}, M., {Vesperini}, E., {et~al.} 2019, \apjl, 884,
  L24

\bibitem[{{Dalessandro} {et~al.}(2018){Dalessandro}, {Cadelano}, {Vesperini},
  {Salaris}, {Ferraro}, {Lanzoni}, {Raso}, {Hong}, {Webb}, \&
  {Zocchi}}]{dalessandro_2018}
{Dalessandro}, E., {Cadelano}, M., {Vesperini}, E., {et~al.} 2018, \apj, 859,
  15

\bibitem[{{Dalessandro} {et~al.}(2021){Dalessandro}, {Raso}, {Kamann},
  {Bellazzini}, {Vesperini}, {Bellini}, \& {Beccari}}]{dalessandro_2021}
{Dalessandro}, E., {Raso}, S., {Kamann}, S., {et~al.} 2021, \mnras, 506, 813

\bibitem[{{Della Croce} {et~al.}(2024){Della Croce}, {Aros}, {Vesperini},
  {Dalessandro}, {Lanzoni}, {Ferraro}, \& {Bhat}}]{della_croce_2024}
{Della Croce}, A., {Aros}, F.~I., {Vesperini}, E., {et~al.} 2024, \aap, 690,
  A179

\bibitem[{{D'Ercole} {et~al.}(2008){D'Ercole}, {Vesperini}, {D'Antona},
  {McMillan}, \& {Recchi}}]{dercole_2008}
{D'Ercole}, A., {Vesperini}, E., {D'Antona}, F., {McMillan}, S. L.~W., \&
  {Recchi}, S. 2008, \mnras, 391, 825

\bibitem[{{Foreman-Mackey} {et~al.}(2013){Foreman-Mackey}, {Hogg}, {Lang}, \&
  {Goodman}}]{emcee_2013}
{Foreman-Mackey}, D., {Hogg}, D.~W., {Lang}, D., \& {Goodman}, J. 2013, \pasp,
  125, 306

\bibitem[{{Fregeau} {et~al.}(2004){Fregeau}, {Cheung}, {Portegies Zwart}, \&
  {Rasio}}]{fregeau_2004}
{Fregeau}, J.~M., {Cheung}, P., {Portegies Zwart}, S.~F., \& {Rasio}, F.~A.
  2004, \mnras, 352, 1

\bibitem[{{Gavagnin} {et~al.}(2016){Gavagnin}, {Mapelli}, \&
  {Lake}}]{gavagnin_2016}
{Gavagnin}, E., {Mapelli}, M., \& {Lake}, G. 2016, \mnras, 461, 1276

\bibitem[{{Gieles} {et~al.}(2018){Gieles}, {Charbonnel}, {Krause},
  {H{\'e}nault-Brunet}, {Agertz}, {Lamers}, {Bastian}, {Gualandris}, {Zocchi},
  \& {Petts}}]{gieles_2018}
{Gieles}, M., {Charbonnel}, C., {Krause}, M. G.~H., {et~al.} 2018, \mnras, 478,
  2461

\bibitem[{{Gieles} \& {Zocchi}(2015)}]{gieles_2015}
{Gieles}, M. \& {Zocchi}, A. 2015, \mnras, 454, 576

\bibitem[{{Gieles} \& {Zocchi}(2017)}]{gieles_2017}
{Gieles}, M. \& {Zocchi}, A. 2017, {LIMEPY: Lowered Isothermal Model Explorer
  in PYthon}, Astrophysics Source Code Library, record ascl:1710.023

\bibitem[{{Giersz} \& {Heggie}(1996)}]{giersz_1996}
{Giersz}, M. \& {Heggie}, D.~C. 1996, \mnras, 279, 1037

\bibitem[{{Giersz} {et~al.}(2013){Giersz}, {Heggie}, {Hurley}, \&
  {Hypki}}]{giersz_2013}
{Giersz}, M., {Heggie}, D.~C., {Hurley}, J.~R., \& {Hypki}, A. 2013, \mnras,
  431, 2184

\bibitem[{{Gratton} {et~al.}(2019){Gratton}, {Bragaglia}, {Carretta},
  {D'Orazi}, {Lucatello}, \& {Sollima}}]{gratton_2019}
{Gratton}, R., {Bragaglia}, A., {Carretta}, E., {et~al.} 2019, \aapr, 27, 8

\bibitem[{{Gratton} {et~al.}(2012){Gratton}, {Carretta}, \&
  {Bragaglia}}]{gratton_2012}
{Gratton}, R.~G., {Carretta}, E., \& {Bragaglia}, A. 2012, \aapr, 20, 50

\bibitem[{Harris {et~al.}(2020)Harris, Millman, van~der Walt, Gommers,
  Virtanen, Cournapeau, Wieser, Taylor, Berg, Smith, Kern, Picus, Hoyer, van
  Kerkwijk, Brett, Haldane, del R{\'{i}}o, Wiebe, Peterson,
  G{\'{e}}rard-Marchant, Sheppard, Reddy, Weckesser, Abbasi, Gohlke, \&
  Oliphant}]{numpy_2020}
Harris, C.~R., Millman, K.~J., van~der Walt, S.~J., {et~al.} 2020, Nature, 585,
  357

\bibitem[{{H{\'e}nault-Brunet} {et~al.}(2015){H{\'e}nault-Brunet}, {Gieles},
  {Agertz}, \& {Read}}]{henault-brunet_2015}
{H{\'e}nault-Brunet}, V., {Gieles}, M., {Agertz}, O., \& {Read}, J.~I. 2015,
  \mnras, 450, 1164

\bibitem[{{H{\'e}non}(1971{\natexlab{a}})}]{henon_1971b}
{H{\'e}non}, M. 1971{\natexlab{a}}, \apss, 13, 284

\bibitem[{{H{\'e}non}(1971{\natexlab{b}})}]{henon_1971a}
{H{\'e}non}, M.~H. 1971{\natexlab{b}}, \apss, 14, 151

\bibitem[{{Hunter}(2007)}]{matplotlib_2007}
{Hunter}, J.~D. 2007, Computing in Science and Engineering, 9, 90

\bibitem[{{Hurley} {et~al.}(2000){Hurley}, {Pols}, \& {Tout}}]{hurley_2000}
{Hurley}, J.~R., {Pols}, O.~R., \& {Tout}, C.~A. 2000, \mnras, 315, 543

\bibitem[{{Hurley} {et~al.}(2002){Hurley}, {Tout}, \& {Pols}}]{hurley_2002}
{Hurley}, J.~R., {Tout}, C.~A., \& {Pols}, O.~R. 2002, \mnras, 329, 897

\bibitem[{{Hypki} \& {Giersz}(2013)}]{hypki_2013}
{Hypki}, A. \& {Giersz}, M. 2013, \mnras, 429, 1221

\bibitem[{{Hypki} {et~al.}(2022){Hypki}, {Giersz}, {Hong}, {Leveque}, {Askar},
  {Belloni}, \& {Otulakowska-Hypka}}]{hypki_2022}
{Hypki}, A., {Giersz}, M., {Hong}, J., {et~al.} 2022, \mnras, 517, 4768

\bibitem[{{Hypki} {et~al.}(2025){Hypki}, {Vesperini}, {Giersz}, {Hong},
  {Askar}, {Otulakowska-Hypka}, {Hellstrom}, \& {Wiktorowicz}}]{hypki_2024}
{Hypki}, A., {Vesperini}, E., {Giersz}, M., {et~al.} 2025, \aap, 693, A41

\bibitem[{{Ishchenko} {et~al.}(2023){Ishchenko}, {Sobolenko}, {Berczik},
  {Omarov}, {Sobodar}, {Kalambay}, \& {Yurin}}]{ishchenko_2023}
{Ishchenko}, M., {Sobolenko}, M., {Berczik}, P., {et~al.} 2023, \aap, 678, A69

\bibitem[{{Khoperskov} {et~al.}(2018){Khoperskov}, {Mastrobuono-Battisti}, {Di
  Matteo}, \& {Haywood}}]{khoperskov_2018}
{Khoperskov}, S., {Mastrobuono-Battisti}, A., {Di Matteo}, P., \& {Haywood}, M.
  2018, \aap, 620, A154

\bibitem[{{King}(1966)}]{king_1966}
{King}, I.~R. 1966, \aj, 71, 64

\bibitem[{{Kroupa}(2001)}]{kroupa_2001}
{Kroupa}, P. 2001, \mnras, 322, 231

\bibitem[{{Lacchin} {et~al.}(2022){Lacchin}, {Calura}, {Vesperini}, \&
  {Mastrobuono-Battisti}}]{lacchin_2022}
{Lacchin}, E., {Calura}, F., {Vesperini}, E., \& {Mastrobuono-Battisti}, A.
  2022, \mnras, 517, 1171

\bibitem[{{Lanzoni} {et~al.}(2016){Lanzoni}, {Ferraro}, {Alessandrini},
  {Dalessandro}, {Vesperini}, \& {Raso}}]{lanzoni_2016}
{Lanzoni}, B., {Ferraro}, F.~R., {Alessandrini}, E., {et~al.} 2016, \apjl, 833,
  L29

\bibitem[{{Leitinger} {et~al.}(2023){Leitinger}, {Baumgardt}, {Cabrera-Ziri},
  {Hilker}, \& {Pancino}}]{leitinger_2023}
{Leitinger}, E., {Baumgardt}, H., {Cabrera-Ziri}, I., {Hilker}, M., \&
  {Pancino}, E. 2023, \mnras, 520, 1456

\bibitem[{{Leitinger} {et~al.}(2025){Leitinger}, {Baumgardt}, {Cabrera-Ziri},
  {Hilker}, {Carbajo-Hijarrubia}, {Gieles}, {Husser}, \&
  {Kamann}}]{leitinger_2025}
{Leitinger}, E.~I., {Baumgardt}, H., {Cabrera-Ziri}, I., {et~al.} 2025, \aap,
  694, A184

\bibitem[{{Libralato} {et~al.}(2022){Libralato}, {Bellini}, {Vesperini},
  {Piotto}, {Milone}, {van der Marel}, {Anderson}, {Aparicio}, {Barbuy},
  {Bedin}, {Borsato}, {Cassisi}, {Dalessandro}, {Ferraro}, {King}, {Lanzoni},
  {Nardiello}, {Ortolani}, {Sarajedini}, \& {Sohn}}]{libralato_2022}
{Libralato}, M., {Bellini}, A., {Vesperini}, E., {et~al.} 2022, \apj, 934, 150

\bibitem[{{Libralato} {et~al.}(2023){Libralato}, {Vesperini}, {Bellini},
  {Milone}, {van der Marel}, {Piotto}, {Anderson}, {Aparicio}, {Barbuy},
  {Bedin}, {Brown}, {Cassisi}, {Nardiello}, {Sarajedini}, \&
  {Scalco}}]{libralato_2023}
{Libralato}, M., {Vesperini}, E., {Bellini}, A., {et~al.} 2023, \apj, 944, 58

\bibitem[{{Livernois} {et~al.}(2024){Livernois}, {Aros}, {Vesperini}, {Askar},
  {Bellini}, {Giersz}, {Hong}, {Hypki}, {Libralato}, \&
  {Ziliotto}}]{livernois_2024}
{Livernois}, A.~R., {Aros}, F.~I., {Vesperini}, E., {et~al.} 2024, \mnras, 534,
  2397

\bibitem[{{Martens} {et~al.}(2023){Martens}, {Kamann}, {Dreizler},
  {G{\"o}ttgens}, {Husser}, {Latour}, {Balakina}, {Krajnovi{\'c}}, {Pechetti},
  \& {Weilbacher}}]{martens_2023}
{Martens}, S., {Kamann}, S., {Dreizler}, S., {et~al.} 2023, \aap, 671, A106

\bibitem[{{Mastrobuono-Battisti} \& {Perets}(2013)}]{mastrobuono-battisti_2013}
{Mastrobuono-Battisti}, A. \& {Perets}, H.~B. 2013, \apj, 779, 85

\bibitem[{{Merritt}(1985)}]{merrit_1985}
{Merritt}, D. 1985, \aj, 90, 1027

\bibitem[{{Milone} \& {Marino}(2022)}]{milone_2022}
{Milone}, A.~P. \& {Marino}, A.~F. 2022, Universe, 8, 359

\bibitem[{{Osipkov}(1979)}]{osipkov_1979}
{Osipkov}, L.~P. 1979, Pisma v Astronomicheskii Zhurnal, 5, 77

\bibitem[{{Pavl{\'\i}k} \& {Vesperini}(2021)}]{pavlik_2021}
{Pavl{\'\i}k}, V. \& {Vesperini}, E. 2021, \mnras, 504, L12

\bibitem[{{Richer} {et~al.}(2013){Richer}, {Heyl}, {Anderson}, {Kalirai},
  {Shara}, {Dotter}, {Fahlman}, \& {Rich}}]{richer_2013}
{Richer}, H.~B., {Heyl}, J., {Anderson}, J., {et~al.} 2013, \apjl, 771, L15

\bibitem[{{Sollima}(2021)}]{sollima_2021}
{Sollima}, A. 2021, \mnras, 502, 1974

\bibitem[{{Spitzer}(1987)}]{spitzer_1987}
{Spitzer}, L. 1987, {Dynamical evolution of globular clusters} (Princeton,
  N.J.: Princeton University Press)

\bibitem[{{Tiongco} {et~al.}(2016){Tiongco}, {Vesperini}, \&
  {Varri}}]{tiongco_2016}
{Tiongco}, M.~A., {Vesperini}, E., \& {Varri}, A.~L. 2016, \mnras, 455, 3693

\bibitem[{{Tiongco} {et~al.}(2019){Tiongco}, {Vesperini}, \&
  {Varri}}]{tiongco_2019}
{Tiongco}, M.~A., {Vesperini}, E., \& {Varri}, A.~L. 2019, \mnras, 487, 5535

\bibitem[{{Vesperini} {et~al.}(2021){Vesperini}, {Hong}, {Giersz}, \&
  {Hypki}}]{vesperini_2021}
{Vesperini}, E., {Hong}, J., {Giersz}, M., \& {Hypki}, A. 2021, \mnras, 502,
  4290

\bibitem[{{Vesperini} {et~al.}(2018){Vesperini}, {Hong}, {Webb}, {D'Antona}, \&
  {D'Ercole}}]{vesperini_2018}
{Vesperini}, E., {Hong}, J., {Webb}, J.~J., {D'Antona}, F., \& {D'Ercole}, A.
  2018, \mnras, 476, 2731

\bibitem[{{Vesperini} {et~al.}(2013){Vesperini}, {McMillan}, {D'Antona}, \&
  {D'Ercole}}]{vesperini_2013}
{Vesperini}, E., {McMillan}, S. L.~W., {D'Antona}, F., \& {D'Ercole}, A. 2013,
  \mnras, 429, 1913

\bibitem[{Virtanen {et~al.}(2020)Virtanen, Gommers, Oliphant, Haberland, Reddy,
  Cournapeau, Burovski, Peterson, Weckesser, Bright, {van der Walt}, Brett,
  Wilson, Millman, Mayorov, Nelson, Jones, Kern, Larson, Carey, Polat, Feng,
  Moore, {VanderPlas}, Laxalde, Perktold, Cimrman, Henriksen, Quintero, Harris,
  Archibald, Ribeiro, Pedregosa, {van Mulbregt}, \& {SciPy 1.0
  Contributors}}]{scipy_2020}
Virtanen, P., Gommers, R., Oliphant, T.~E., {et~al.} 2020, Nature Methods, 17,
  261

\bibitem[{{Wang} {et~al.}(2020){Wang}, {Kroupa}, {Takahashi}, \&
  {Jerabkova}}]{wang_2020}
{Wang}, L., {Kroupa}, P., {Takahashi}, K., \& {Jerabkova}, T. 2020, \mnras,
  491, 440

\bibitem[{{Watkins} {et~al.}(2022){Watkins}, {van der Marel}, {Libralato},
  {Bellini}, {Anderson}, \& {Alfaro-Cuello}}]{watkins_2022}
{Watkins}, L.~L., {van der Marel}, R.~P., {Libralato}, M., {et~al.} 2022, \apj,
  936, 154

\bibitem[{{Weatherford} {et~al.}(2018){Weatherford}, {Chatterjee}, {Rodriguez},
  \& {Rasio}}]{weatherford_2018}
{Weatherford}, N.~C., {Chatterjee}, S., {Rodriguez}, C.~L., \& {Rasio}, F.~A.
  2018, \apj, 864, 13

\end{thebibliography}

\begin{appendix}

\onecolumn
\section{Simulation properties at $7\,\text{Gyr}$ and  $12\,\text{Gyr}$}

\begin{table*}[h!]
\caption{\label{tab_app:sim_prop_t07} Properties at $t=7\,\text{Gyr}$}
\centering
\begin{tabular}{l|c|c|c|c|l}
\hline\hline
Name  & $M_{\text{tot}}$ ($10^6\,M_{\odot}$) & $M_{\text{tot}}/M_{\text{tot,0}}$ & $N_{\text{tot}}/N_{\text{tot,0}}$ & $M_{\text{2P}}/M_{\text{tot}}$ & $R_{\text{hl}}\,(\text{pc})$ \\
\hline
     N2M &     0.25 &     0.20 &     0.34 &     0.52 &     4.31 \\ 
     N1M &     0.11 &     0.17 &     0.29 &     0.54 &     3.13 \\ 
  N1M-ra &     0.13 &     0.21 &     0.35 &     0.46 &     3.80 \\ 
    N05M &     0.04 &     0.13 &     0.21 &     0.56 &     1.77 \\ 
 N05M-wf &     0.06 &     0.19 &     0.32 &     0.50 &     3.91 \\
 \hline
 N2M-r10 &     0.26 &     0.21 &     0.36 &     0.48 &     5.61 \\ 
 N1M-r10 &     0.12 &     0.19 &     0.33 &     0.50 &     4.19 
\\  
\hline
\end{tabular}
\tablefoot{$M_{\text{tot}}$ is the total mass of the cluster at $7\,\text{Gyr}$, $M_{\text{tot}}/M_{\text{tot,0}}$ is the ratio of the total mass at $7\,\text{Gyr}$ to the total initial mass, $N_{\text{tot}}/N_{\text{tot,0}}$ is the ratio of the total number of stars at $7\,\text{Gyr}$ to the total initial number of stars. $M_{\text{2P}}/M_{\text{tot}}$ is the ratio of the mass in 2P stars, $M_{\text{2P}}$, to the total clusters' mass, $M_{\text{tot}}$ at $7\,\text{Gyr}$. $R_{\text{hl}}$ is the cluster's half-light radius.}
\end{table*}

\begin{table*}[h!]
\caption{\label{tab_app:sim_prop_t12} Properties at $t=12\,\text{Gyr}$}
\centering
\begin{tabular}{l|c|c|c|c|l}
\hline\hline
Name  & $M_{\text{tot}}$ ($10^6\,M_{\odot}$) & $M_{\text{tot}}/M_{\text{tot,0}}$ & $N_{\text{tot}}/N_{\text{tot,0}}$ & $M_{\text{2P}}/M_{\text{tot}}$ & $R_{\text{hl}}\,(\text{pc})$ \\
\hline
     N2M &     0.20 &     0.16 &     0.27 &     0.58 &     3.66 \\ 
     N1M &     0.08 &     0.12 &     0.20 &     0.60 &     3.48 \\ 
  N1M-ra &     0.10 &     0.16 &     0.26 &     0.52 &     3.12 \\ 
    N05M &     0.02 &     0.07 &     0.09 &     0.61 &     1.65 \\ 
 N05M-wf &     0.04 &     0.14 &     0.23 &     0.57 &     3.34 \\
 \hline
 N2M-r10 &     0.22 &     0.17 &     0.30 &     0.54 &     4.33 \\ 
 N1M-r10 &     0.09 &     0.15 &     0.25 &     0.55 &     3.33  
\\  
\hline
\end{tabular}
\tablefoot{Same as in Table \ref{tab_app:sim_prop_t07}, but for properties at $12\,\text{Gyr}$.}
\end{table*}

\twocolumn

\onecolumn
\section{Additional Figures}
\begin{figure*}[h!]
    \centering
    \includegraphics[width=\figwidth\linewidth]{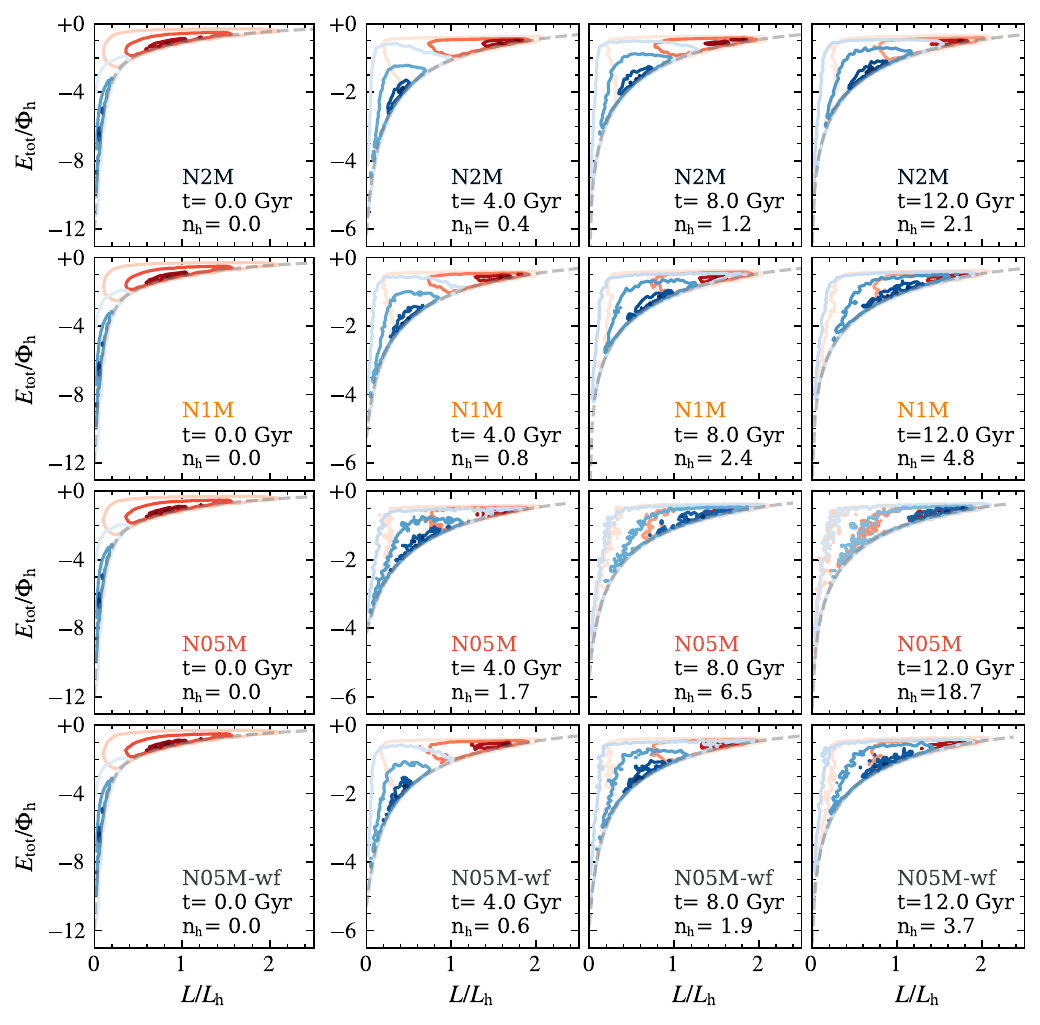}
    \caption{Phase-space distribution evolution for models N2M, N1M, N05M and N05M-wf at \tml{0}{Gyr}, \tml{4}{Gyr}, \tml{7}{Gyr} and \tml{12}{Gyr}. For each model, we include the corresponding dynamical age traced by the number of half-mass relaxation times $n_{\text{h}}$.}
    \label{fig-app:phase-space-all-A}
\end{figure*}

\begin{figure*}[h!]
    \centering
    \includegraphics[width=\figwidth\linewidth]{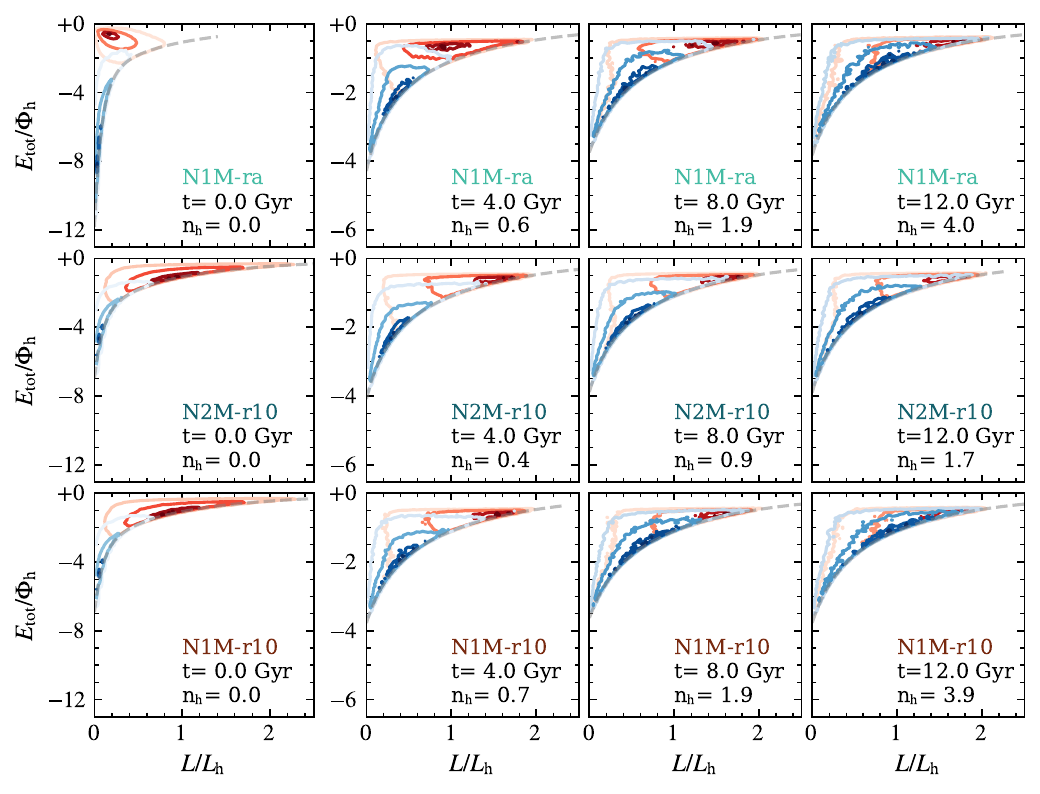}
    \caption{As in Figure \protect\ref{fig-app:phase-space-all-A}, but for models N1M-ra, N2M-r10 and N1M-r10.}
    \label{fig-app:phase-space-all-B}
\end{figure*}

\begin{figure*}[h!]
    \centering
    \includegraphics[width=\figwidth\linewidth]{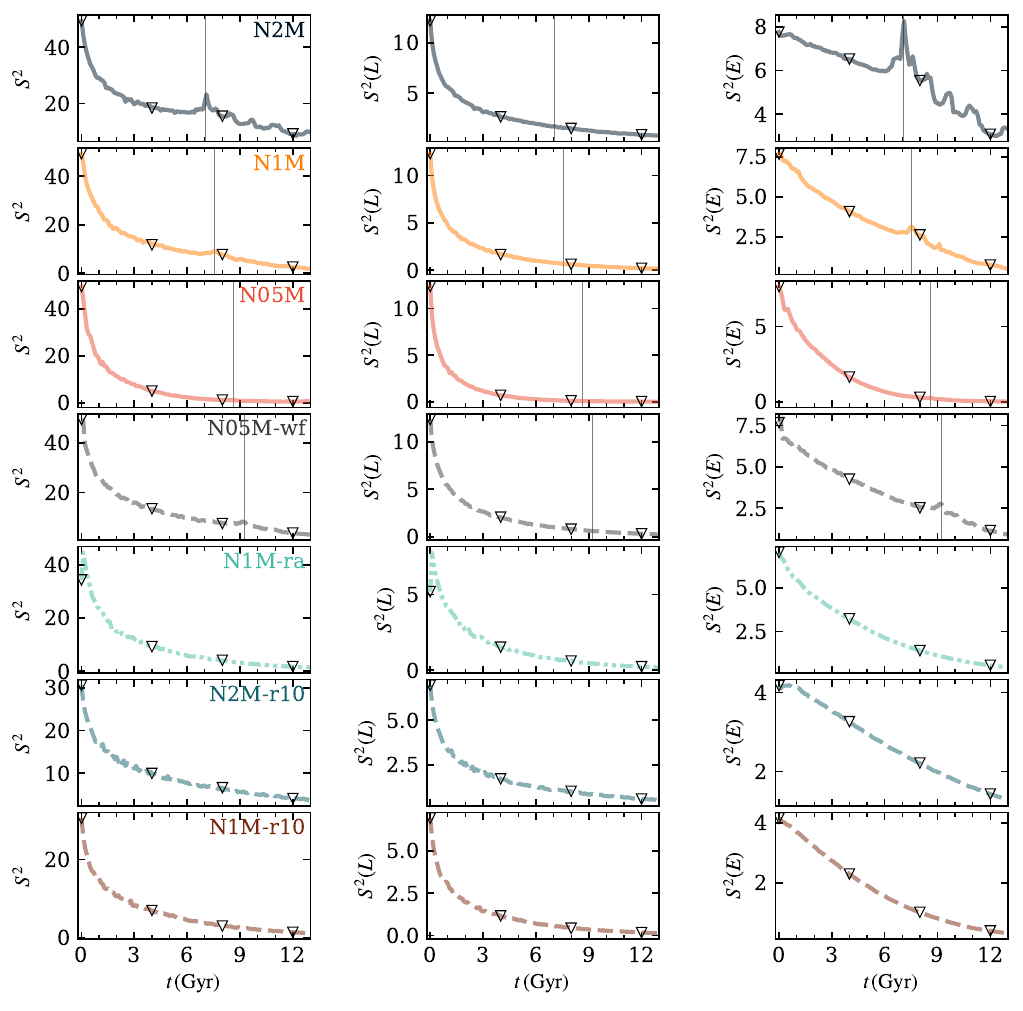}
    \caption{Time evolution of the $S^2$ parameter for all models as in panel (a) in Figure \ref{fig:s2-comparison}, but separated and also including the marginalisation over angular momentum $S^2(L)$ and energy $S^2(E)$. We have marked a vertical line at the time of the first core-collapse for all models that undergo core-collapse within the \tms{13}{Gyr} of evolution.}
    \label{fig-app:all_diff_3d}
\end{figure*}

We include in this appendix three additional figures to complement Figures \ref{fig:phase-space-mixing-example} and \ref{fig:s2-comparison}. Figures \ref{fig-app:phase-space-all-A} and \ref{fig-app:phase-space-all-B} show the phase space distribution for all models at $t={ 0, 4, 8, 12}\,\text{Gyr}$, while Figure \ref{fig-app:all_diff_3d} shows the $S^2$ parameter (as in Figure \ref{fig:s2-comparison}) and the corresponding marginalised version for angular momentum $S^2(L)$ and energy $S^2(E)$.

\end{appendix}
\end{document}